\documentclass[aps,twocolumn,prd,superscriptaddress,longbibliography]{revtex4-1}

\usepackage[utf8]{inputenc}
\usepackage{csquotes}

\usepackage{mathtools}
\usepackage{amsfonts}
\usepackage{mathrsfs}
\usepackage{bm}
\usepackage{listings}
\usepackage{graphicx}
\usepackage{color}
\usepackage[dvipsnames]{xcolor}
\usepackage{array}
\usepackage[abs]{overpic}

\usepackage{placeins}

\usepackage{makecell}
\usepackage{subcaption}

\usepackage{xspace}
\usepackage{siunitx}
\usepackage{xfrac}
\usepackage{hyperref}
\usepackage[nameinlink]{cleveref}
\usepackage{appendix}

\usepackage{xifthen}
\usepackage{xcolor}
\hypersetup{
	colorlinks,
	linkcolor={red!75!black},
	citecolor={blue!75!black},
	urlcolor={blue!75!black}
}

\usepackage{booktabs}
\usepackage{multirow}

\newcolumntype{C}{>{$}c<{$}}
\AtBeginDocument{
	\heavyrulewidth=.08em
	\lightrulewidth=.05em
	\cmidrulewidth=.03em
	\belowrulesep=.65ex
	\belowbottomsep=0pt
	\aboverulesep=.4ex
	\abovetopsep=0pt
	\cmidrulesep=\doublerulesep
	\cmidrulekern=.5em
	\defaultaddspace=.5em

}

\newcommand{\tr}{{\text{tr}}}

\renewcommand{\Re}{\operatorname{Re}}
\renewcommand{\Im}{\operatorname{Im}}

\newcolumntype{L}{>{\centering\arraybackslash}m{3cm}}

\newenvironment{eq}{
    \begin{equation}
    \begin{aligned}
}{
    \end{aligned}
    \end{equation}
    \ignorespacesafterend
}

\graphicspath{{./figures/}}

\newcommand{\gettitle}{Quantum Systems from Random Probabilistic Automata}

\hypersetup{
	pdftitle={\gettitle},
	pdfauthor={Kreuzkamp, Wetterich},
	pdfkeywords=
	{Quantum Mechanics} {Cellular Automaton},
	bookmarksopen=true,
	bookmarksopenlevel=2,
	bookmarksnumbered=true
}

\begin{document}

\title{\gettitle}

\author{A. Kreuzkamp}
\author{C. Wetterich}

\affiliation{Institut für Theoretische Physik, Universität Heidelberg, Philosophenweg 16, 69120 Heidelberg, Germany}

\begin{abstract}
Probabilistic cellular automata with deterministic updating are quantum systems.
We employ the quantum formalism for an investigation of random probabilistic cellular automata, which start with a probability distribution over initial configurations. The properties of the deterministic updating are randomly distributed over space and time. We are interested in a possible continuum limit for a very large number of cells. As an example we consider bits with two colors, moving to the left or right on a linear chain.
At randomly distributed scattering points, they change direction and color.
A numerical simulation reveals the typical features of quantum systems. We find particular initial probability distributions which reemerge periodically after a certain number of time steps, as produced by the periodic evolution of energy eigenstates in quantum mechanics.
Using a description in terms of wave functions allows to introduce statistical observables for momentum and energy.
They characterize the probabilistic information without taking definite values for a given bit configuration, with a conceptual status similar to temperature in classical statistical thermal equilibrium.
Conservation of energy and momentum are essential ingredients for the understanding of the evolution of our stochastic probabilistic automata.
This evolution resembles in some aspects a single Dirac fermion in two dimensions with a random potential.
\end{abstract}

\maketitle

\section{Introduction} \label{sect:introduction}

Cellular automata \cite{JVN, ULA, ZUS} have found applications in wide areas of science \cite{GAR, LIRO, TOOM, DKT, WOLF, VICH, PREDU, TOMA, FLN, HED, RICH, AMPA, HPP, CREU}.
While the cells as basic building blocks and the updating steps of an automaton can be very simple, rather complex dynamics can emerge after many updating steps. Focusing on invertible automata, our aim is the understanding of the behavior for a very large number of cells after many time steps. In particular, we are interested in a possible continuum limit for which important simplifications may occur. An automaton can be described by an updating rule how a configuration of $N$ bits at $t$ is mapped to a new bit-configuration at $t+\epsilon$.
For very large $N$ the number of possible bit-configurations $2^N$ grows huge and only a probabilistic setting seems meaningful. As $N$ increases, the numerical simulations that we perform in this note rapidly encounter practical limitations.
For the investigation of a possible continuum limit one needs to combine simulations with an analytic understanding which could be extrapolated to the limit $N \rightarrow \infty$. We propose here to use the formalism of quantum mechanics for the analytic description.
We are not aware of any other methods for this purpose which work for the case where the automaton is too complex for allowing direct combinatorial solutions.

For invertible automata no information is lost by the updating. This type of deterministic evolution can be described by a unitary step evolution operator \cite{CWIT, CWQF, CWPW2020}. In turn, unitary matrices can be represented in terms of a Hermitian Hamiltonian. The Hamiltonian description of the evolution is used by t'Hooft for his interesting proposal of a deterministic interpretation of quantum mechanics based on selected \textquote{ontological} observables \cite{GTH, ELZE, HOOFT2, HOOFT3, HOOFT4}.
In contrast, for our approach the probabilistic setting will be crucial. It is implemented by specifying at some initial time $t_{in}$ a probability distribution for the $2^N$ configurations.
We associate to each configuration $\tau$ a probability $w_\tau(t_{in})$ and investigate how the probability distribution $\{w_\tau(t)\}$ evolves with time as a consequence of the updating. While the updating rule remains deterministic, the probabilistic aspects enter by the initial state. The \textquote{probabilistic automata} defined in this way are \textquote{classical statistical systems} based solely on the standard axioms for probabilities, without any additional input. All quantum properties will follow from this.

We do not consider a probabilistic updating, for which the names of probabilistic or stochastic automata are used as well \cite{AAR, GTH, GJH, LMS, PA, FU, MM, RLK}.
For a probabilistic updating one deals with Markov chains for which the long time behavior typically (but not always) approaches some equilibrium state.
Probabilistic updatings have strong connections to equilibrium statistical systems \cite{VER, PE, DK, DOM, RU, GD}.

The cells of our automaton are labeled by $N_x$ discrete points $x$ on a one-dimensional chain. The updating of a given cell $x$ is only influenced by the bit-configurations of the neighboring cells at $x \pm \epsilon$. This property of cellular automata induces a causal structure with \textquote{light cones}, as familiar from particle physics.
We may actually identify the bits with fermions in an occupation number basis. The configurations at a given time $t$ are given by bits or \textquote{occupation numbers} $n_\gamma(t, x)$ that can take the values one or zero.
For $n_\gamma(t,x) = 1$ a fermion of type $\gamma$ is present at the position $x$ at the time $t$, while for $n_\gamma(t, x) = 0$ it is absent. The possible bit-configurations $\{n_\gamma(t, x)\}$ at a given $t$ correspond to the possible basis states of a multi-fermion quantum system. (For a more profound investigation of the correspondence between probabilistic cellular automata and fermionic quantum field theories and the corresponding general map to a functional integral for Grassmann-variables see ref. \cite{CWFCS, CWFGI, CWPCA, CWFCB, CWNEW, CWFPPCA}.)

We aim here for a system that remains simple enough to allow for numerical simulations, and complex enough such that methods beyond explicit combinatorial solutions become necessary. We choose a system of four species of bits $\gamma = 1...4$, distinguished by two colors, red and green, and the property of being right- or left-movers.
For free fermions the right movers move at each updating step one position to the right , from $x$ to $x + \epsilon$, while left-movers move to the left.
The free propagation is modified by scattering points $\overline{x}_i(t)$. When a fermion encounters a scattering point it changes direction and color.

A certain number of scattering points is distributed randomly at every $t$ on positions $\overline{x}_i(t)$ on the chain. Each such distribution defines a different automaton, which we call \textquote{random automaton} in view of the randomly chosen distribution. We consider, however, fixed distributions and do not consider averages over distributions of scattering points. Due to the irregularity of the randomly chosen distribution of scattering points a combinatorial treatment becomes rapidly very involved as the number of cells and the number of scattering points increases. We implement a certain amount of regularity in time by repeating the same distribution of scattering points after time intervals $\Delta t$ which comprise a fixed number of time steps $\epsilon$, equal to the lattice distance $\epsilon$. This implements time translation invariance by \textquote{mesoscopic time steps} $\Delta t$.

\begin{figure}[t]\centering
    \includegraphics[width=8.5cm]{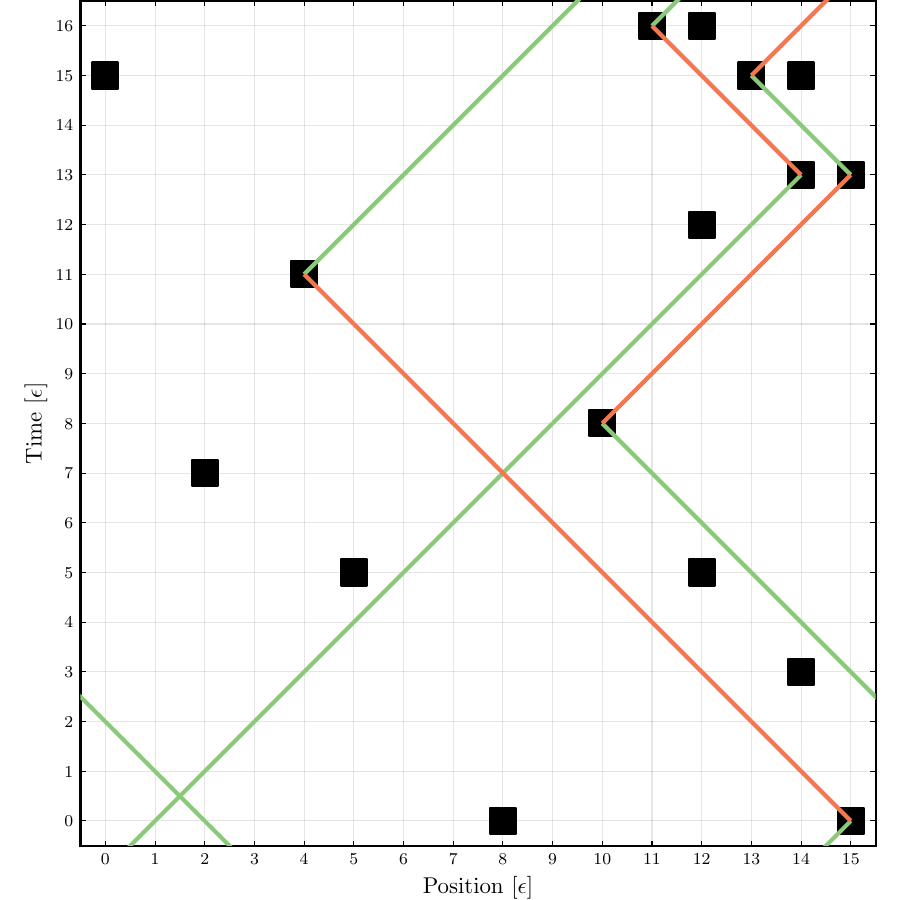}
    \caption{Three possible trajectories of single particles. At scattering points denoted by black squares the particles change direction and color. The (fixed) scattering points have been chosen randomly.}
    \label{fig:ism_example_trajectories}
\end{figure}

We restrict the discussion in this note to the very simple configurations where only a single bit is occupied and all others are empty. These configurations can be labeled by the position $x$ and the type $\gamma$ of the single fermion present. The probability distributions for these single-bit configurations are labeled by $w_\gamma(t, x) \geq 0, \quad \sum_\gamma \sum_x w_\gamma(t, x) = 1$. Our updating rule conserves the total number of occupied bits, such that we can use this type of probability distribution for all $t$.
A numerical investigation follows the trajectory $(x, \gamma)(t)$ for the single occupied bit according to the updating rule.
The probability for each \textquote{point} on the trajectory is the same as for the point of the trajectory $(x, \gamma)(t_{in})$ at the initial time $t_{in}$. In this way we can construct $w_\gamma(t,x)$ for arbitrary initial $w_\gamma(t_{in}, x)$.
We display in fig. \labelcref{fig:ism_example_trajectories} three trajectories, for a distribution of scattering points shown as black squares.

A single-particle state in a fermionic quantum field theory is much more involved, being an excitation of a complex half-filled vacuum state \cite{CWPW2020}. Our single-bit state offers the advantage that the number of relevant configurations is reduced from $2^{4N_x}$ to $4N_x$ and therefore amenable to numerical studies. The random scattering points could be interpreted as mimicking a non-trivial vacuum \cite{CWPCAQP}. The one-bit stochastic automata form quantum systems for which the Hamiltonian is not known explicitly. They therefore offer a good starting point for an investigation how quantum properties characterize the evolution of general probabilistic automata.

\begin{figure}[t]\centering
    \includegraphics[width=8.5cm]{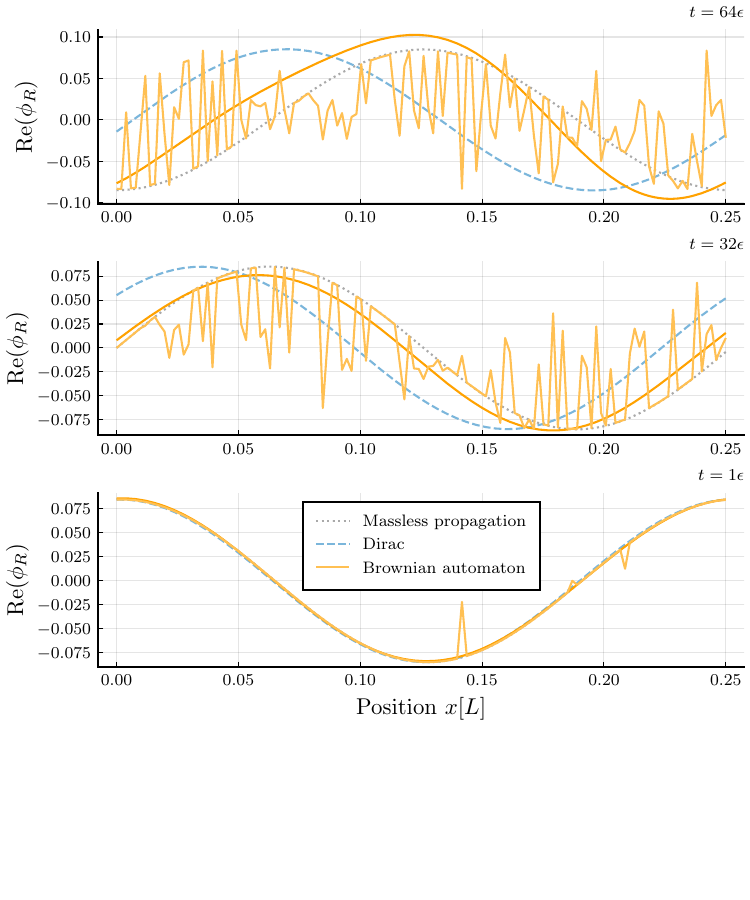}
    \caption{Evolution of the wave function for a Brownian automaton, model A. Parameters are given by eq. \labelcref{eq:brownianModelParams}. We start with a plane wave solution of the Dirac equation and display the result of the updating at three different times. Due to the scattering at randomly distributed scattering points the originally smooth wave function becomes rough. The smoothened orange curve only keeps a few lowest Fourier modes of the distribution in space. This may be compared with the absence of scattering points (gray dotted curve), for which the automaton is exactly equivalent to a free massless Dirac particle. We also compare to the evolution of a massive Dirac particle (blue curve).}
    \label{fig:brownianModel_timeEvolution}
\end{figure}

For a moderate number of points on the chain with length $L$, $N_x = L / \epsilon$, we can follow the evolution of a given initial probability  distribution numerically by following trajectories for all $4N_x$ initial configurations.
The result for a particular random probabilistic automaton, also called \textquote{Brownian automaton}, to be specified later (\textquote{model A}), is shown in fig. \labelcref{fig:brownianModel_timeEvolution}. We parametrize here the probability distribution for a given type $\gamma$ by a \textquote{real wave function} $q_\gamma(t,x)$ with probabilities $w_\gamma(t, x) = q_\gamma^2(t, x)$.
We start at $t_{in}=0$ with a smooth probability distribution for which $q_\gamma(t_{in}, x)$ are simple harmonics, corresponding to a solution of the Dirac equation with mass $m = 2.5 \cdot 2\pi / L$ and momentum $p = 4 \cdot 2\pi / L$. Fig. \labelcref{fig:brownianModel_timeEvolution} displays $q_1(t, x)$ after a certain number of time steps. At $t=\epsilon$ only the bits at a few initial positions have encountered a scattering point, while for most initial positions no scattering has happened and the fermion has moved one position to the right. The scattered bits result in the figure by the small local deviations from the harmonics for the unscattered bits. For $t=16\epsilon$ a plane wave for free fermions would have moved by $16$ positions to the right. However, now a larger number of bits have scattered, and the probability distribution shows sizeable fluctuations in space.
Averaging over short distance fluctuations one can still perceive a harmonics. The extrema of the harmonics have moved to the right, as expected for a wave with positive velocity. However, the harmonics is now displaced somewhat to the left as compared to a model without scattering. Due to scattering the average velocity of the motion is reduced. We also compare this with the evolution of the wave function for a Dirac fermion with the same mass and momentum as used for the initial wave function of the Brownian automaton. As time progresses further the probability distribution seems to become more and more random, as shown for $t = 64 \epsilon$.

\begin{figure}[t]\centering
    \includegraphics[width=8.5cm]{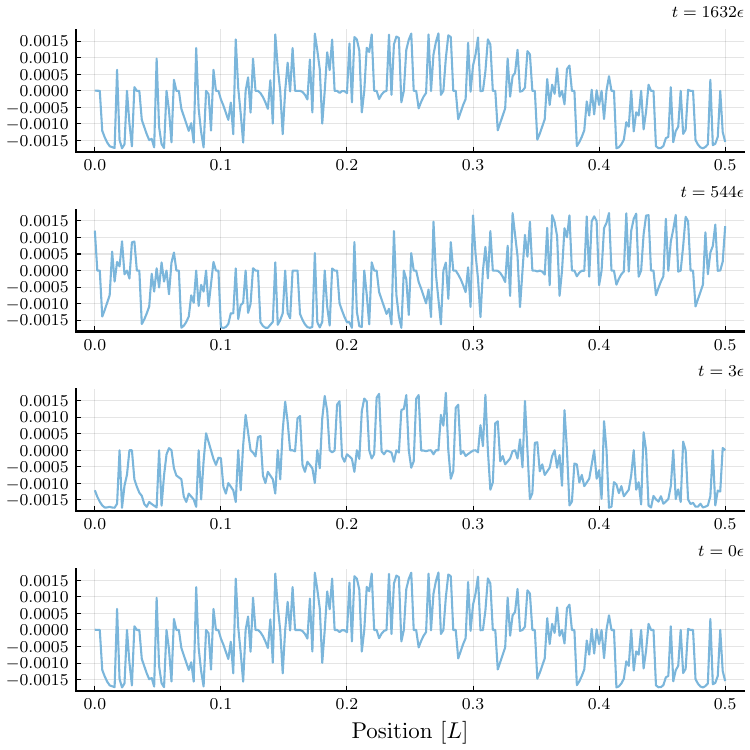}
    \caption{Time evolution of a stationary state, evolved with a periodic random PCA, model B. Parameters are given by eq. \labelcref{eq:pRPCA_params}. We display the difference of mean occupation numbers $\langle n_1(t, x) - n_3(t, x) \rangle = w_1(t, x) - w_3(t,x)$ after different numbers of time steps. The precise reappearance of the initial distribution (bottom figure) after $1632$ time steps (top figure) is striking.}
    \label{fig:ism_qm_red_occupationNum_timeEvolution}
\end{figure}

An evolution towards randomness is however, not the fate of all initial probability distributions. There exist particular initial probability distributions which first seem to move towards randomness as time progresses, but are then returned periodically after a certain number of time steps. Quantum mechanics tells us that this is a genuine feature, but in practice it is not easy to find the initial probability distributions which correspond to energy eigenstates. Only for rather simple systems we have been able to construct them explicitly. For a periodic random scattering automaton (\textquote{model B}), a periodic distribution is shown in fig. \labelcref{fig:ism_qm_red_occupationNum_timeEvolution} where we display the difference of mean occupation numbers $\langle n_1(t, x) - n_3(t, x) \rangle = w_1(t, x) - w_3(t,x)$ at different times. The initial form of the distribution of the difference of mean occupation numbers is displayed in the bottom figure. After three time steps the form of the distribution has changed substantially. Instead of getting more and more random or chaotic, the initial distribution is recovered at each integer multiple of $17$ time steps, albeit displaced to the right. After $1632$ time steps one finds precisely the initial distribution.
The periodicity in time of this probability distribution is characteristic for an energy eigenstate in quantum mechanics.

Quite generically, the time evolution of classical statistical systems can be described by a generalized quantum formalism \cite{CWIT, CWQF} with wave functions and operators for observables. For probabilistic automata the time evolution is unitary. In this case the generalized quantum formalism reduces to the standard quantum formalism.
In the presence of  suitable complex structure the real wave function $q_\gamma (t, x)$ can be encoded in a complex wave function $\psi_\alpha (t,x)$. In our case the two colors are associated to the real and imaginary parts of $\psi$.
Probabilistic automata are then quantum systems in the usual complex formulation \cite{CWPW2020}. Based on this insight we will be able to construct for our random automaton observables for momentum and energy, as familiar for single-particle quantum mechanics.
Momentum and energy conservation will be the key for the understanding of many features of the dynamics of our random cellular automata.

While our random probabilistic automata are definitely quantum systems, there is no guarantee that the corresponding Hamiltonian is \textquote{smooth enough} to permit a simple continuum limit.
One may speculate that the roughness of a strong but rare random scattering could be overcome by some averaging over space and time. For establishing a continuum limit one may hope that at least the low energy eigenstates of the Hamiltonian may have eigenfunctions that are smooth in some coarse grained sense. The present paper develops several tools for describing coarse graining in space and time. Again, the full power of the quantum formalism with density matrices and a change of basis is needed for this purpose.
So far, we have not succeeded to construct a smooth continuum limit that remains valid for large time. It remains open if this is due to the limited number of points $N_x$ available for numerical simulations, the practical limitation to find energy eigenstates, or if the quantum system itself does not admit a smooth continuum limit.

We start in section \labelcref{sect:discreteQM} by a short discussion of discrete quantum mechanics. It introduces the concepts and notations that we will employ for the analysis of the random probabilistic automata.
We briefly review one-particle states for Dirac fermions in one space and one time dimension that will be employed for comparison with the quantum description of the random automaton. In sect. \labelcref{sect:randomPCA} we discuss the quantum formalism for cellular automata based on wave functions and a discrete Schrödinger equation. We introduce the particular random probabilistic cellular automata studied in this work.

Section \labelcref{sect:momentum} is devoted to the momentum observable and the associated quantum operator. This type of observable may at first sight not be expected for the automaton, while it is a basic notion for quantum systems of particles.
The momentum observable does not have a fixed value for a given bit-configuration $\{n_\gamma(t,x)\}$. It is rather characterizing properties of the probability distribution or wave function and belongs therefore to a class of \textquote{statistical observables} without sharp values in the \textquote{microstates} of the statistical ensemble. For such observables Bell's inequalities \cite{BELL, CHSH} for classical statistical correlations do not apply\cite{CWPW2020}.
The momentum observable requires the probabilistic setting for the automaton. The associated momentum operator may have a somewhat complex form in the position basis.
Its simplicity becomes apparent in the momentum basis after a Fourier transform.
Again, only the quantum formalism enables the powerful tool of basis transformations for cellular automata.

In section \labelcref{sect:energy} we turn to the energy observable and the associated Hamilton operator, which exists due to the regularity implemented through a periodicity of scattering points after a time interval $\Delta t$. The discrete time translation symmetry by mesoscopic time intervals $\Delta t$ induces the corresponding conservation law, as known from quantum mechanics.
The Hamiltonian $H$ is therefore defined on the mesoscopic level. Arbitrary functions of $H$ are conserved quantities.
We discuss eigenstates of the Hamiltonian and the corresponding periodicity in time, which we have demonstrated in fig. \labelcref{fig:ism_qm_red_occupationNum_timeEvolution}. In this section we also make contact to combinatorial constructions of energy eigenstates for cases where the number of scattering points is not too large. \textquote{Single-orbit states} are \textquote{probabilistic clocks} \cite{CWPW2020} and suitable states show periodicity for the time evolution of the probability distribution.
Quantum mechanics allows superpositions of wave functions. The superposition of two different eigenfunctions to a given energy eigenvalue defines again an energy eigenstate.
This construction, which typically becomes important in the continuum limit, needs the wave function for the description of the probabilistic information. It cannot be implemented on the level of probability distributions.

In sect. \labelcref{sect:densityMatrixAndCoarseGraining} we enter the road towards the continuum limit. We introduce for probabilistic automata the density matrix of quantum mechanics, which allows well known coarse graining by taking subtraces. We both discuss subtraces in position and momentum space. For systems which are invariant under space translations by $\Delta x$ we establish a conserved coarse grained momentum observable. The associated operator commutes with the Hamiltonian. The coarse grained momentum is therefore a conserved quantity. Simultaneous eigenstates of coarse grained momentum and energy are very useful for the understanding of the dynamics of the random probabilistic automata.
In sect. \labelcref{sect:coarseGrainedEvolution} we address possible notions of coarse graining in time. In contrast to coarse graining in space this cannot proceed by a coarse grained density matrix. A possible road could be the focus on an effective model for small energy eigenvalues, somewhat similar to concepts in particle physics.

Sect. \labelcref{sect:continuumLimit} is devoted to the continuum limit. Any continuum limit requires a sufficient smoothness of the wave function, at least on some coarse grained level. We therefore perform an investigation of the fate of smooth initial wave functions, typically plane waves which are solutions to a corresponding Dirac fermion. As long as the wave function remains smooth enough we can perform a \textquote{naive continuum limit}. In this naive continuum limit the Hamiltonian of the random probabilistic cellular automaton coincides with the one for a massive Dirac fermion, with mass given by the mean number of scattering points. The numerical simulation finds for the initial stages of the evolution indeed many aspects of the one for the Dirac fermion.
Sect. \labelcref{sect:discussion} contains our conclusions.

\section{Discrete quantum mechanics} \label{sect:discreteQM}
In this section, we briefly describe a discretization of quantum mechanics for a single particle. For this purpose, space points are put on a discrete lattice, and the evolution is described by discrete time steps. There are no new concepts here. Some type of discretization is usually done for any numerical solution of the Schrödinger equation. We put discrete quantum mechanics into a form that can be used directly for probabilistic cellular automata.

\subsection{Discretization in space}
We consider a particle in one dimension. Space points $x$ are put in equal distance $\epsilon$ on a circle with length \( L \) by using periodic boundary conditions. For a finite number \( N_x \) of space points, the wave function \(\psi_\gamma(t, x)\) belongs to a finite-dimensional Hilbert space. With an internal index \( \gamma \) for different particle species taking \( M \) values, the wave function at a given \( t \) is specified by \( M N_x \) complex numbers. The usual infinite-dimensional Hilbert space obtains in the limit \( \epsilon \rightarrow 0 \) at fixed \( L \), or \( N_x \rightarrow \infty \).

\subsection{Discrete time evolution}
The time evolution for a discrete time step \( \epsilon \) is specified by a unitary step evolution operator
\begin{eq}\label{eq:unitaryStepOperator}
 U(t) = U(t+\epsilon, t), \quad \psi(t+\epsilon) = U(t) \psi(t).
\end{eq}
This replaces the continuous Schrödinger equation. The step evolution operator is related to the continuum Hamiltonian \( H^{(c)}(t) \) of a continuous formulation by
\begin{eq}
U(t) = \exp\left(-i \int_{t}^{t+\epsilon} dt' H^{(c)}(t') \right) = \exp(-i \epsilon \overline{H}(t)).
\end{eq}

Inversely, the continuum formulation is recovered in the limit \( \epsilon \rightarrow 0 \),
\begin{eq}
\frac{i}{\epsilon} [\psi(t+\epsilon) - \psi(t)] &= \frac{1}{\epsilon} \int_{t}^{t+\epsilon} dt' H^{(c)}(t') \psi(t) \\
    &= \overline{H}(t) \psi(t).
\end{eq}
For differentiable \( \psi(t) \) the l.h.s. of the last equation reads \( i \partial_t \psi(t) \). We will consider settings where \( U(t) \) or \( \overline{H}(t) \) depend on time. For discrete space the derivatives \( \partial_x \) appearing in \( \overline{H} \) transfer to suitable lattice derivatives. Besides being a matrix in position space, both \( U \) and \( \overline{H} \) are also matrices in internal space. We often omit internal indices or the space labels if the meaning is clear.

Quantum mechanics admits a real formulation with a real wave function \( q(t) \) with a doubled number of components,
\begin{eq}
q(t) = \begin{pmatrix}q_r(t) \\ q_i(t)\end{pmatrix} \, , \quad \psi(t) = q_r(t) + i q_i(t).
\end{eq}
The real functions \( q_r \) and \( q_i \), correspond to the real and imaginary part of the complex wave function \( \psi \).
This real formulation helps to build the bridge to the classical statistical formulation of cellular automata.
In the real formulation, the step evolution operator \( \hat{S}(t) \) is a real orthogonal matrix, \( U = U_r + i U_i \),
\begin{eq}\label{eq:realStepEvolutionOperator}
q(t+\epsilon) = \hat{S}(t) q(t), \quad \hat{S} = \begin{pmatrix}U_r & -U_i \\ U_i & U_r\end{pmatrix}.
\end{eq}
Inversely, one can reformulate a real evolution equation \labelcref{eq:realStepEvolutionOperator} as a complex wave equation \labelcref{eq:unitaryStepOperator} provided \( \hat{S}(t) \) is compatible with a suitable complex structure.
Besides the choice \labelcref{eq:realStepEvolutionOperator}, different complex structures are possible, see the discussion in \cite{CWPW2020}.

\subsection{Mesoscopic evolution operator}

A certain number of time steps may be grouped into a mesoscopic time step \( \Delta t \). Here \( \Delta t \) may be much larger than \( \epsilon \), but still small as compared to observable macroscopic time steps. The mesoscopic evolution operator \( \overline{U}(t) \) is defined as the sequence of step evolution operators
\begin{eq}\label{eq:mesoscopicEvolutionOperator}
\overline{U}(t) = &U(t+\Delta t-\epsilon) U(t+\Delta t-2\epsilon)\\
                  &\cdots U(t+\epsilon)U(t), \\
\psi(t + \Delta t) = &\overline{U}(t) \psi(t).
\end{eq}
It is guaranteed to be unitary and hence can be expressed in terms of a hermitian Hamiltonian \( H(t) \),
\begin{eq} \label{eq:hermitianHamiltonian}
\overline{U}^\dagger(t) \overline{U}(t) &= 1,\\
\overline{U}(t) &= \exp(-i \Delta t H(t)), \\
H^\dagger(t) &= H(t).
\end{eq}
The relation \labelcref{eq:hermitianHamiltonian} defines the (mesoscopic) Hamiltonian \( H(t) \) which will be a central concept for our investigation.

The unitarity of \( \overline{U} \) implies that \( H(t) \) is hermitian. If \( \psi(t) \) is sufficiently smooth on time scales \( \Delta t \) we can again infer from the discrete time evolution equation \labelcref{eq:mesoscopicEvolutionOperator} a Schrödinger equation, now with the mesoscopic Hamiltonian \( H(t) \). We will focus on a setting where \( H(t) \) is independent of \( t \). This means that the sequence of step evolution operators $U$ \labelcref{eq:mesoscopicEvolutionOperator} is repeated identically after a number of steps corresponding to \( \Delta t \). The system exhibits discrete time-translation invariance by steps $\Delta t$. For a constant mesoscopic Hamiltonian \( H \) the Schrödinger equation takes the standard form
\begin{eq}
i \partial_t \psi(t) = H \psi(t).
\end{eq}
A solution of this differential equation with initial value given by \( \psi(0) \) coincides with the solution of the discrete evolution equation \labelcref{eq:unitaryStepOperator} for all discrete time points \( t = n \Delta t \), with integer \( n \).

\subsection{Single free Dirac particle in two dimensions}

In one space and one time dimension the complex wave function of a single free Dirac fermion with mass $m$ obeys the Dirac equation ( \( \partial_0 = \partial_t, \partial_1 = \partial_x, \mu = (0, 1) \), summation over repeated indices always implied),
\begin{eq}
 \gamma^\mu \partial_\mu \psi + m \psi = 0, \quad \psi = \begin{pmatrix} \psi_{R} \\ \psi_{L} \end{pmatrix}.
\end{eq}
Here \( \psi \) has two complex components and we choose a real representation of the Dirac matrices
\begin{eq}
\gamma^0 = -i \tau_2 = \begin{pmatrix} 0 & -1 \\ 1 & 0 \end{pmatrix}, \quad \gamma^1 = \tau_1 = \begin{pmatrix} 0 & 1 \\ 1 & 0 \end{pmatrix}.
\end{eq}
The corresponding continuous Schrödinger equation,
\begin{eq} \label{eq:continuumSchroedinger}
i \partial_t \psi = H^{(c)} \psi, \quad H^{(c)} = -i \partial_x \tau_3  + m \tau_2,
\end{eq}
does not mix the real and imaginary parts of $\psi$
\begin{eq}
(\partial_t + \partial_x) \psi_R &= - m \psi_L, \\
(\partial_t - \partial_x) \psi_L &= + m \psi_R. \\
\end{eq}
The step evolution operator \( U(t) \) is therefore a real orthogonal \( 2 \times 2 \) matrix in internal space. For the corresponding real formulation of quantum mechanics, \( \hat{S}(t) \) is a real \( 4 \times 4 \) matrix in internal space, with \( U_i = 0 \) in eq. \labelcref{eq:realStepEvolutionOperator}.

For a massless particle, \( m=0 \), the upper component \( \psi_R \) describes a right-moving particle with general solution
\begin{eq}
\psi_R(t, x) = f_R(t-x).
\end{eq}
After a time step \( \epsilon \) the wave function is displaced in space by \( \epsilon \) in the positive \( x \)-direction. Correspondingly, \( \psi_L \) is a left mover. For \( m=0 \), the step evolution operator is a real block diagonal matrix (we omit the time argument, since $H^{(c)}$ does not depend on $t$),
\begin{eq} \label{eq:discreteFreeDiracEvolutionInTwoDim}
U_{f} &=
\begin{pmatrix}
U_R & 0 \\
0 & U_L
\end{pmatrix},\\
U_L(x,x') = \delta_{x,x'-\epsilon}&, \quad U_R(x,x') = \delta_{x,x'+\epsilon},
\end{eq}
realizing
\begin{eq}
\psi_R(t+\epsilon, x) &= \sum_{x'} U_R(x,x') \psi_R(t,x') \\
    &= \psi_R(t, x-\epsilon).
\end{eq}
The corresponding free Hamiltonian \( \overline{H}_{f} \) is defined for the discrete setting by
\begin{eq} \label{eq:freeStepOperator}
U_{f} = \exp(-i\epsilon \overline{H}_{f}).
\end{eq}
It involves a suitable lattice derivative \( \partial_x \) as expected \cite{CWPCAQP}.

For \( m \neq 0 \) we take for the step evolution operator the product
\begin{eq} \label{eq:mesoscopicStepOperator}
U &= U_{m} U_{f},\\
U_{m}(x,x') &= \exp(-i \epsilon \overline{H}_{m}) \delta_{x,x'} = \exp(-i \epsilon m \tau_2) \delta_{x,x'} \\
&=
\begin{pmatrix}
\cos(\epsilon m) & -\sin(\epsilon m) \\
\sin(\epsilon m) & \cos(\epsilon m)
\end{pmatrix} \delta_{x,x'}.
\end{eq}
The step evolution operator $U$ for the Dirac particle
is a real orthogonal matrix. It does not mix the real and imaginary
parts of $\psi$ which therefore evolve independently. We can associate
these independent parts with Majorana fermions.

The corresponding discrete Hamiltonian \( \overline{H} \) obeys
\begin{eq}
U &= e^{-i\epsilon \overline{H}} = e^{-i\epsilon \overline{H}_{m}} e^{-i\epsilon \overline{H}_{f}} \\
    &= e^{-i\epsilon(\overline{H}_{m} + \overline{H}_{f})} + O(\epsilon^2 [\overline{H}_{f}, \overline{H}_{m}]).
\end{eq}
For wave functions that are sufficiently smooth on the scale \( \epsilon \), the commutator term \( \sim\epsilon^2 \) becomes negligible and the continuum Hamiltonian \labelcref{eq:continuumSchroedinger} is recovered. For fixed $\Delta t$ and different $\epsilon$  we have compared the solution of the discrete evolution with step evolution operator \labelcref{eq:mesoscopicStepOperator} with a leap frog integration of the continuous Schrödinger equation \labelcref{eq:continuumSchroedinger} or analytic solutions. For $\epsilon m$ in the order of magnitude used for the remainder of this paper, we found good agreement and therefore an acceptable continuum limit for this discretization of the Dirac equation.

The physical properties of the system do not change if we subtract from $H$ a constant piece $m$. The corresponding continuum Hamiltonian,
\begin{eq} \label{eq:nonrealHamiltonian}
\tilde{H}^{(c)} = -i \tau_3 \partial_x + m(\tau_2 - 1),
\end{eq}
is no longer purely imaginary, such that the evolution mixes now real and imaginary parts of the wave function. Correspondingly, in the discrete formulation $U_m$ is multiplied by a phase
\begin{eq} \label{eq:phaseMultipliedStepEvolution}
    \tilde{U}_m(x, x') = e^{i \epsilon m} U_m(x, x').
\end{eq}
In this version a space-independent wave function,
\begin{eq} \label{eq:spaceIndependentWavefunction}
    \psi(x) = \frac{1}{\sqrt{2N_x}} \begin{pmatrix} 1 \\ i \end{pmatrix},
\end{eq}
does not change in time, $\tilde{U}_m \psi = \psi, \quad U_f \psi = \psi$.

\section{Random probabilistic cellular automata} \label{sect:randomPCA}

An automaton is defined by a deterministic updating of a \textquote{state} \( \rho \) to a new state \( \overline{\tau}(\rho) \). More precisely, the updating maps a bit-configuration \( \rho \) at time \( t \) to a unique new configuration \( \tau \) at time \( t+\epsilon \). We consider invertible automata for which the map \( \overline{\tau}(\rho) \) is invertible. As outlined in the introduction, we specify for single-bit configurations \( \tau \) or \( \rho \) by a discrete coordinate \( x \) and an internal index \( \gamma = 1...4 \), \( \tau = (x, \gamma) \). They denote position and type of the single \textquote{occupied} bit or fermion.

Probabilistic automata are characterized by a probability distribution over initial configurations. For a random probabilistic automaton the prescription for the updating steps involves elements that are chosen partly randomly. Nevertheless, a given random automaton has fixed updating steps such that the evolution of a given initial configuration remains deterministic.

\subsection{Step evolution operator and wave function for probabilistic automata}

The updating rule can be expressed in terms of a step evolution operator \( \hat{S}(t) \) acting on a real wave function \( q_\gamma(t,x) \),
\begin{eq} \label{eq:automatonEvolution}
q_\tau(t+\epsilon) &= \hat{S}_{\tau\rho}(t) q_\rho(t),\\
q_\gamma(t+\epsilon,x) &= \sum_{x'} \hat{S}_{\gamma\delta}(t;x,x')q_{\delta}(t,x').
\end{eq}
This step evolution operator has to be a \textquote{unique jump matrix} for which in each row and column a single element takes the value \( \pm 1 \), and all other elements are zero (no sum here):
\begin{eq} \label{eq:uniqueJumpMatrix}
\hat{S}_{\tau \rho} = \sigma_{\tau} \delta_{\tau, \overline{\tau}(\rho)}, \quad \text{with } \sigma_{\tau} = \pm 1.
\end{eq}
Here \( \overline{\tau}(\rho) \) encodes the updating map.

A deterministic automaton (e.g. a deterministic computer) is at initial time $t=0$ in a definite configuration $\rho_{in}$. This initial configuration is characterized by a sharp wave function with a single non-zero component
\begin{eq}
    q_\rho(0) = \pm \delta_{\rho, \rho_{in}}.
\end{eq}
At time \( \epsilon \) the evolution \labelcref{eq:automatonEvolution}, \labelcref{eq:uniqueJumpMatrix} yields again a sharp wave function
\begin{eq}
q_{\tau}(\epsilon) = \pm \delta_{\tau, \overline{\tau}(\rho_{in})},
\end{eq}
with unique non-zero element \( \overline{\tau}(\rho_{in}) \) corresponding to the updated configuration \( \rho_{in} \). This continues for further updating steps such that at \( t = n\epsilon \), the nonzero component \( q_\tau(n\epsilon) \) corresponds to the sequence of updatings of the initial configuration \( \rho_{in} \).
The different updating steps need not be identical, such that \( \hat{S}(t) \) can depend on time.

A probabilistic automaton is characterized by a probability distribution over the possible initial configurations \( w_\rho(0) \). The update remains deterministic, such that
\begin{eq}\label{eq:probabilityUpdating}
 w_{\tau}(\epsilon) = w_{\overline{\rho}(\tau)}(0),
\end{eq}
The probability for the configuration \( \tau \) at \( t=\epsilon \) is precisely the probability for the configuration \( \overline{\rho}(\tau) \) at \( t=0 \) from which \( \tau \) has originated by the updating. This continues to further time steps. The probabilities \( w_\tau(t) \) are expressed in terms of the real \textquote{classical} wave function \cite{CWQPCS, wetterich_classical_2011} as
\begin{eq} \label{eq:probabilityAndRealWavefunction}
w_{\tau}(t) = q_{\tau}^2(t).
\end{eq}
The evolution law \labelcref{eq:automatonEvolution}, \labelcref{eq:uniqueJumpMatrix} yields
\begin{eq}
    q_{\tau}(\epsilon) = \pm q_{\overline{\rho}(\tau)}(0),
\end{eq}
and therefore accounts for the updating law \labelcref{eq:probabilityUpdating} of the probability distribution.

The use of the wave function is a redundant description since the sign of \( q_{\tau} \) does not affect the probability \( w_{\tau} \). The freedom in the choice of signs for \( q_{\tau} \) corresponds to a local discrete gauge symmetry. The evolution law remains invariant if a change of signs in the wave function is accompanied by a corresponding change of signs in the step evolution operator. A given sign convention for the step evolution operator can be considered as a gauge fixing. Observables do not depend on the choice of signs\cite{CWPW2020}.

There is no additional physical information in the signs of $q_\tau$. Nevertheless, the use of the wave function offers several important advantages.
First, the updating corresponds to a rotation of the unit vector $q$.
This guarantees the normalization of the probability distribution. Eq. \labelcref{eq:probabilityAndRealWavefunction} guarantees positive probabilities, $w_\tau(t) \geq 0$.
Second, from the wave function one can construct a density matrix and apply the coarse graining procedures of quantum mechanics.
Third, and most important for our purpose, the formulation in terms of a wave function allows us to apply the full formalism of quantum mechanics to probabilistic automata.
In particular, the linear evolution law \labelcref{eq:automatonEvolution} implies the superposition principle of quantum mechanics.

\subsection{Probabilistic cellular automata}

For a cellular automaton the updating of the configuration  of a given cell only depends on the configuration of a few neighboring cells. In our context we identify the cells with the positions $x$. For a cellular automaton the step evolution $\hat{S}_{\gamma\delta}(x, x')$ differs from zero only for $x'$ in the neighborhood of $x$ (including $x'=x$). The cellular property implies a causal structure and the concept of (generalized) light cones.

The propagation part of the step evolution operator for the Dirac fermion, $U_f$ in eq. \labelcref{eq:discreteFreeDiracEvolutionInTwoDim} is already a unique jump matrix. For the corresponding cellular automaton one has two species of right-movers and two species of left-movers, denoted as red and green. A complex structure is easily introduced by encoding $q_\gamma(t,x)$ in a two-component complex wave function $\psi(t,x)$,
\begin{eq}
    \psi = \begin{pmatrix}\psi_R \\ \psi_L\end{pmatrix},\quad
    \psi_R = q_1 + i q_2, \quad \psi_L = q_3 + i q_4.
\end{eq}
The phase of the complex wave function encodes the relative probability of finding a red or green particle, with an invariance under complex conjugation and negation, both of which do not alter the individual probabilities $w_{1} = \Re(\psi_R)^2, w_{2} = \Im(\psi_R)^2$.

In the complex language, the step evolution operator for this automaton is given by a $2 \times 2$ matrix in internal space.
For $\hat{S}_f$ transporting $q_1$ and $q_2$ one position to the right, and $q_3$ and $q_4$ to the left, the corresponding matrix $U_f$ is given by eq. \labelcref{eq:discreteFreeDiracEvolutionInTwoDim}.
For $U_m=1$, the probabilistic automaton describes precisely the time evolution of a free massless Dirac particle.
In contrast, the part $U_m$ for the Dirac particle, eq. \labelcref{eq:mesoscopicStepOperator} or \labelcref{eq:phaseMultipliedStepEvolution}, is not a unique jump matrix for non-zero $\epsilon m \ll 1$. For the automata discussed in this paper, we have to replace $U_m$ by a unique jump matrix. More precisely, we consider a structure similar to eq. \labelcref{eq:mesoscopicStepOperator}
\begin{eq} \label{eq:TwoStepTimeEvolution}
 U(t) = U_s(t)U_f,
\end{eq}
with $U_s(t)$ a unique jump matrix which may depend now on $t$. For the \textquote{scattering operator} $U_s(t)$, we take a local structure given in the complex picture by
\begin{eq}
 U_{s,\alpha\beta}(t; x, x') = W_{\alpha\beta}(t,x) \delta_{x,x'}.
\end{eq}
The $2 \times 2$ matrices $W(t,x)$ are either given by $\eta\tau_2$ or by the unit matrix. The choices $\eta = \pm i$ or $\eta = \pm 1$ ensure the unique jump property.

For $\eta = -i$ the matrix $U_s$ is real. Similar to the Majorana basis for a Dirac fermion the real and imaginary parts of $\psi$ evolve independently. Nevertheless, following simultaneously the evolution of both parts of the wave function will allow us to employ the complex formulation for a simple implementation of the Fourier transform. In contrast, for $\eta = 1$ the matrix $U_s$ is purely imaginary and therefore mixes real and imaginary parts of $\psi$. In the real formulation one easily verifies the unique jump property
\begin{eq} \label{eq:uniqueJumpMatrixEtaOne}
    q = 
    \begin{pmatrix}
    q_{Rr} \\ q_{Ri} \\ q_{Lr} \\ q_{Li}
    \end{pmatrix}
    = 
    \begin{pmatrix}
    q_1 \\ q_2 \\ q_3 \\ q_4
    \end{pmatrix},
    \quad
    \hat{S}_m =
    \begin{pmatrix}
    0 & 0 & 0 & 1 \\
    0 & 0 &-1 & 0 \\
    0 &-1 & 0 & 0 \\
    1 & 0 & 0 & 0 \\
    \end{pmatrix}
    \delta_{x, x'}.
\end{eq}
The choice $\eta = 1$ is closer to eq. \labelcref{eq:nonrealHamiltonian}. Indeed, the space-independent wave function \labelcref{eq:spaceIndependentWavefunction} does not change with time. In contrast to the choice $\eta=\pm i$ it is an eigenstate of the Hamiltonian with eigenvalue zero. For this reason we concentrate on $\eta=1$ in the following.

The internal part of the matrix $\hat{S}_m$ in eq. \labelcref{eq:uniqueJumpMatrixEtaOne} has two eigenvalues $+1$ and two eigenvalues $-1$. The linearly independent eigenfunctions for the eigenvalue $+1$ are $q_4 = q_1, q_2=q_3=0$, corresponding to eq. \labelcref{eq:spaceIndependentWavefunction}, and $q_3=-q_2, q_1=q_4=0$, which multiplies eq. \labelcref{eq:spaceIndependentWavefunction} by $i$. For wave functions close to these eigenfunctions and with only a small variation in space, one expects that $U_s(t)U_f$ results only in a small change of the wave function. This may be an interesting starting point for a continuum limit.

\subsection{Random probabilistic cellular automata}

We consider a spacetime region $(\Delta t, \Delta x)$ of space and time points in the intervals $0 \leq x < \Delta x, 0 \leq t < \Delta t$. Within this region, we distribute a certain number of scattering points $(\overline{t}_j, \overline{x}_j)$. For any scattering point, we take $W(\overline{t}_j, \overline{x}_j) = \eta\tau_2$, and choose $W(t, x) = 1$ otherwise.
The combined step evolution operator $U(t)$ in eq. \labelcref{eq:TwoStepTimeEvolution} is still a unique jump matrix, such that we describe an automaton. The updating of the cell $x$ only involves the cells $x-\epsilon$ and $x+\epsilon$, which ensures the cellular property.

At every scattering point, a right-mover is scattered into a left-mover and vice versa, whereas without a scattering point, the particle continues its motion. The idea is that this occasional scattering somehow mimics effects of a mass term which likewise switches between right-movers and left-movers. We take this pattern periodic in $x$, with period $\Delta x$, and periodic in $t$, with period $\Delta t$.
(For convenience we may shift the boundaries of the intervals keeping the number of sites in the interval, or $\Delta x$ and $\Delta t$ fixed.)
A given cellular automaton is then completely defined by the distribution of scattering points in the interval $[\Delta t, \Delta x]$.
If we specialize to $\Delta x = L$ (keeping in mind periodicity in $x$) we have to specify for the $\Delta t$-interval the distribution of scattering points over the  whole range of $x$.
Each distribution defines a quantum system with a mesoscopic Hamiltonian $H$, defined by eqs. \labelcref{eq:mesoscopicEvolutionOperator}, \labelcref{eq:hermitianHamiltonian}. Indeed, each $U(t)$ is a unitary matrix, such that $\overline{U}$ is unitary as well. This guarantees $H^\dagger = H$. The periodicity in time makes $H$ independent of $t$. Different distributions of scattering points define different quantum systems with different Hamiltonians $H$.

The distribution of scattering points in the interval $(\Delta t, \Delta x)$ is kept fixed.
Intuitively, rare scattering points may be considered as the analogue of a small mass. In the absence of scattering points, we recover the automaton that precisely describes the quantum system of a free massless Dirac fermion.
In the other extreme, if every point in the interval is a scattering point, all right-movers are turned to left-movers at every time step and vice versa.
As a result, the wave function is the same after two time steps, such that for $\Delta t$ comprising an even number of time steps, one has $H=0$.
This rather trivial automaton does not describe a propagating particle.
If a large fraction of points in $\Delta x$ are scattering points, one does not expect a behavior close to a Dirac particle.

If the total number of scattering points $n_{tot}$ in the interval $(\Delta t, \Delta x)$ is much smaller than $\Delta x / \epsilon$, within any time interval $\Delta t$ most trajectories of particle positions do not involve a single scattering, and therefore remain as straight lines.
The mesoscopic Hamiltonian of this type of automaton is expected to deviate again strongly from the one for a massive Dirac fermion.
On the other hand, one may envisage large $\Delta t$ with only a small mean number of scatterings at any point $x$, at a given $t$.
The mean number of scattering points at a given $t$ reads $\hat{n} = n_{tot} \epsilon / \Delta t$, and the mean number per site obeys $\overline{n} = \hat{n}(t) \epsilon / \Delta x$.
For $\overline{n} \ll 1$ the rare scattering may correspond to small $m \epsilon$, while the total number of scatterings in the interval $(\Delta t, \Delta x)$, namely $n_{tot} = \Delta t  \Delta x \overline{n} / \epsilon^2$, can be much larger than $\Delta x / \epsilon$, such that almost every particle trajectory undergoes at least one scattering in every time interval $\Delta t$.
One may ask if such an automaton could mimic certain aspects of a massive Dirac fermion.
We recall, however, that the Dirac particle can be seen as a homogeneous distribution of small scatterings at every point, while the probabilistic automata have maximal scattering at rare points.

The significance of particular space points or time points might be reduced by distributing a large number of scattering points $n_{tot} \gg \Delta x / \epsilon$ randomly in the interval $(\Delta t, \Delta x)$. The corresponding automaton may be called a random probabilistic automaton (RPCA).
We will discuss two types of RPCA. For the \textquote{Brownian automaton} we take $\Delta x = L$, without additional periodicity in the space direction. For the \textquote{periodic random probabilistic automaton} we assume periodicity in the distribution of scattering points by a certain $\Delta x = M_x \epsilon \ll L$ in the $x$-direction. These automata show a discrete space-translation invariance by $\Delta x$.

For a random automaton with large $n_{tot}$ it seems very hard to gain analytic understanding by following explicitly the trajectories of particle positions. Nevertheless, we will see that important insight on many characteristic features of the automaton can be obtained. This builds on the fact that probabilistic automata are quantum systems and uses the power of the quantum formalism. Naively, one could expect that due to the random scattering the probability distribution always reaches for large time a kind of equilibrium state, typically with equal mean occupation numbers for right- and left-movers. This is prevented, however, by the presence of conserved quantities as momentum and energy. These conserved quantities become visible in the quantum formalism.

\section{Momentum observable} \label{sect:momentum}

The momentum is a key observable for the description of quantum particles. We may therefore investigate its role for the RPCA. It may seem rather unfamiliar to use the notion of a momentum observable for a cellular automaton. However, our description of the automaton as a quantum system permits us to employ all the operators of quantum mechanics. This requires the probabilistic setting and the formulation in terms of a complex wave function. We also can exploit the relation between symmetries and conserved quantities, the latter being represented by operators which commute with the Hamiltonian. We will base our definition of the momentum operator on a Fourier transform. Again, the powerful instrument of basis transformations relies on the formulation with a complex wave function. (There exist generalizations to real wave functions without a complex structure \cite{CWPW2020}.)

\subsection{Discrete Fourier Transform}

A complex wave function $\psi(x)$ for discrete lattice points $x$ on a circle with length $L=N_x \epsilon$ can be expanded in terms of its Fourier components $\psi(q)$,
\begin{eq}
 \psi(x) &= N_x^{-\frac{1}{2}} \sum_q \exp(i q x)\psi(q),\\
 \psi(q) &= N_x^{-\frac{1}{2}} \sum_x \exp(-i q x)\psi(x).
\end{eq}
The momenta are discrete and their number equals $N_x$,
\begin{eq} \label{eq:discreteMomenta}
 q = \frac{2\pi k}{\epsilon N_x}, \quad x = \epsilon j,
\end{eq}
with $k$ and $j$ integers in the interval $[-\frac{N_x}{2}, \frac{N_x}{2}]$. (For even $N_x$, the boundaries are identified. We identify $\sum_x$ with the sum over $j$ and $\sum_q$ with the sum over $k$.)

As familiar for lattices, $q$ is periodic, with $k$ and $k + N_x$ identified. Writing $\psi(x) = \psi(j), \psi(q) = \psi(k)$, the Fourier transform corresponds to a basis transformation with the unitary $N_x \times N_x$ matrix $D$,
\begin{eq}
 \psi(j) &= \sum_k D^{-1}(j, k) \psi(k),\\
 D^{-1}(j, k) &= N_x^{-\frac{1}{2}} \exp(\frac{2\pi i}{N_x} j k).
\end{eq}
This employs the identity
\begin{eq} \label{eq:fourierOrthogonality}
 \frac{1}{N_x} \sum_j \exp(\frac{2\pi i}{N_x} j (k -l)) = \tilde{\delta}_{k,l},
\end{eq}
where $\tilde{\delta}_{k,l} = 1$ for $l=k \mod N_x$, and $\tilde{\delta}_{k,l} = 0$ otherwise.

\subsection{Momentum operator}

The momentum operator is defined in the Fourier basis as
\begin{eq}
 P(q, q') = q \tilde{\delta}_{q, q'}, \quad P(k, k') = \frac{2\pi k}{\epsilon N_x} \tilde{\delta}_{k, k'}.
\end{eq}
The expectation value of an arbitrary function of $P$ obeys the quantum rule
\begin{eq} \label{eq:quantumRule}
 \left<{f(P)}\right> &= \sum_{q, q'} \psi^\dagger(q) (f(P))(q, q') \psi(q)\\
    &= \sum_k \psi^\dagger (k) f(\frac{2\pi k}{\epsilon N_x}) \psi(k).
\end{eq}
We can identify the momentum distribution
\begin{eq} \label{eq:momentumDistribution}
 w(q) = \psi^\dagger(q)\psi(q), \quad \left<{f(P)}\right> = \sum_q f(q)w(q),
\end{eq}
where $w(q)$ denotes the probability for a given momentum $q$. In terms of the momentum operator the free part of the step evolution operator \labelcref{eq:freeStepOperator} takes the simple form \cite{CWFPPCA}
\begin{eq} \label{eq:freeStepOperatorMomSpace}
    \overline{H}_f = P \tau_3.
\end{eq}
This underlines the usefulness of the Fourier transform for the cellular automaton description of free particles. The simple form \labelcref{eq:freeStepOperatorMomSpace} and the direct relation to the Fourier transform are the main reason for this choice of the momentum operator (Alternative definitions are based on the lattice derivative \cite{CWFPPCA}.)

One may express the momentum operator in the position basis
\begin{eq} \label{eq:discreteMomOpInPosSpace}
 \tilde{P}(j, j') &= \sum_{k, k'} D^{-1}(j, k) P(k, k') D(k', j')\\
                  &= \sum_{k} \frac{2\pi k}{\epsilon N_x^2} \exp(\frac{2\pi i k (j - j')}{N_x})\\
                  &= \frac{1}{N_x} \sum_q q \exp(i q (x-x')).
\end{eq}
In the continuum limit this becomes the usual expression
\begin{eq} \label{eq:continuumMomOpInPosSpace}
 \tilde{P}(x, x') = -i \partial_x \delta(x-x').
\end{eq}
We will not need the explicit discrete expression \labelcref{eq:discreteMomOpInPosSpace} since it is much simpler to transform first the wave function to the Fourier basis.

%
%
%
%
\subsection{Plane waves and wave packets}

Plane waves are particular solutions of the Dirac equation
\begin{eq} \label{eq:plainWave}
 \psi_p(t,x) = N_x^{-\frac{1}{2}} \exp(ipx - i \sqrt{p^2 + m^2}t)
    \begin{pmatrix} f(p) \\ i f(-p) \end{pmatrix},
\end{eq}
with
\begin{eq} \label{eq:plainWaveParameter}
 f(p) = \left(\frac{1}{2} (1 + \frac{p}{\sqrt{p^2 + m^2}})\right)^{\frac{1}{2}}.
\end{eq}
They are eigenstates to the momentum operator with eigenvalue $p$, and eigenstates of the Hamiltonian with energy $\sqrt{p^2 + m^2}$. For the continuous Dirac equation these are exact solutions. 

One expects similar stationary solutions for our discrete setting for Dirac fermions. For this discrete setting the normalization condition reads
\begin{eq} \label{eq:wavefunctionNormalizationCondition}
 \sum_x \psi_p^\dagger(x) \psi_p(x) = 1.
\end{eq}
The mean occupation number of right-movers is given by
\begin{eq}
 \left< n_R(x) \right> = \psi_R^\dagger (x) \psi_R(x) = \frac{1}{2N_x} \left(1 + \frac{p}{\sqrt{p^2 + m^2}}\right),
\end{eq}
while for the left-movers one switches the sign of $p$. For positive $p$ one observes an imbalance in favor of the right-movers.

The Dirac equation has solutions with positive and negative energy. The solutions with negative energy can be associated with antiparticles. We focus here on particles. By the constant shift in the Hamiltonian \labelcref{eq:nonrealHamiltonian} the plane wave solutions are then given by
\begin{eq}
    \psi_p(t, x) = N_x^{-\frac{1}{2}} \exp\{i p x - i E(p) t\} 
    \begin{pmatrix}
        f(p) \\ i f(-p)
    \end{pmatrix},
\end{eq}
with
\begin{eq}
    E(p) = \sqrt{p^2 + m^2} - m.
\end{eq}
The discrete Fourier transform shows a sharp momentum
\begin{eq}
 \psi_p(q) = \exp(-i E(q) t) \begin{pmatrix}f(q) \\ i f(-q) \end{pmatrix} \tilde{\delta}_{p, q}.
\end{eq}
Wave packets replace the $\tilde{\delta}$-distribution by a smooth function $w_p(q)$, for example a Gaussian centered around $p$.

For the RPCA we can still consider plane waves as momentum eigenstates and consider, for example, initial plane waves \labelcref{eq:plainWave}. These plane waves are no longer eigenstates of the Hamiltonian, however. Initial plane waves will change to different wave functions in the course of the evolution, as visible in fig. \labelcref{fig:brownianModel_timeEvolution}. For a comparison with the Dirac particle one may start with a plane wave at $t=0$ and follow the evolution according to the discrete step evolution operator either for the Dirac particle or the random cellular automaton. As an example, we take the Brownian automaton with parameters
\begin{eq} \label{eq:brownianModelParams}
    N_x = 512, \quad M_t = \frac{\Delta t}{\epsilon} = 16,\\
    m = 2.5 \cdot \frac{2\pi}{L}, \quad p = 4 \cdot \frac{2\pi}{L}.\\
\end{eq}
We call this parameter set \textquote{model A}.
The results of the comparison are displayed in fig. \labelcref{fig:brownianModel_timeEvolution}. For the Dirac particle the wave function remains smooth as $t$ increases, whereas the rare but strong scattering events of the random automaton lead rather fast to a roughening of the wave function.
This may not be surprising since momentum is not conserved and the initial state is a superposition of many energy eigenstates, see later.

\subsection{Momentum conservation}

For a free Dirac particle, momentum is a conserved quantity. In continuous quantum mechanics, this results from the vanishing commutator of the momentum operator with the Hamiltonian $[P, H] = 0$, expressing the fundamental connection between translation symmetry and conserved momentum.
For the discrete setting, one retains the connection between symmetries and conserved quantities. The issue is conveniently formulated in the Heisenberg picture for operators. In the Heisenberg picture, the momentum operator becomes time-dependent
\begin{eq}
 P_H(t) = \overline{U}^{-1}(t) P \overline{U}(t),
\end{eq}
where we take for $t$ multiples of $\Delta t$. For $P_H(\Delta t) = P(0) = P$ the expectation values of arbitrary functions $f(P)$ are the same for all $t = m\Delta t$, $m$ integer.

Let us first consider $\Delta t = \epsilon$,
\begin{eq}
 P_H(\epsilon) = U_f^{-1} U_{int}^{-1} P U_{int} U_f.
\end{eq}
For the Dirac particle, one has $U_{int} = U_m$, c.f. eq. \labelcref{eq:mesoscopicStepOperator}, while for the random cellular automaton, eq. \labelcref{eq:TwoStepTimeEvolution} yields $U_{int} = U_s(0)$. From eq. \labelcref{eq:discreteMomOpInPosSpace} follows directly $\left[U_m, P\right] = 0$. For the free part, we transform $U_f$ to Fourier space
\begin{eq}
 U_f(k, k') &= \sum_{j, j'} D(k, j) U(j, j') D^{-1}(j',k') \\
              &= \frac{1}{N_x} \exp(\frac{2\pi i}{N_x} (j'k' - jk)) U(j, j').
\end{eq}
With
\begin{eq}
 U_f(j, j') = \begin{pmatrix}
               \delta_{j, j'+1} & 0 \\
               0                & \delta_{j, j'-1}
              \end{pmatrix},
\end{eq}
this yields
\begin{eq}
 U_f(q, q') = \begin{pmatrix}
                 \exp(-iq\epsilon) & 0\\
                 0                 & \exp(iq \epsilon)
                \end{pmatrix}
                \tilde{\delta}_{q, q'}
\end{eq}
such that $\left[ U_f, P \right] = 0$. For the Dirac particle, $U$ commutes with $P$ and momentum is therefore conserved. This does not hold for the random automaton, since $U_s(0)$ does not commute with $P$.

For the periodic random automaton, the invariance under translations by $\Delta x$ is reflected by the conservation of momentum modulo $2\pi / \Delta x$. From $\overline{U}(j+M_x, j' + M_x) = U(j, j'),\quad M_x=\Delta x / \epsilon$, one infers in momentum space
\begin{eq} \label{eq:coarseGrainedMomentumConservation}
 \overline{U}(k, k') = \exp(\frac{2\pi i M_x}{N_x} m (k' - k)) \overline{U}(k, k'),
\end{eq}
for any integer $m$ in the interval $\left[-N_x / (2M_x), N_x/(2M_x)\right]$. (We assume an integer number of $\Delta x$-intervals $\overline{N}_x = N_x/M_x$ contained in $L$.) Taking an average of eq. \labelcref{eq:coarseGrainedMomentumConservation} over these intervals yields
\begin{eq}
 \overline{U}(k, k') &= \frac{1}{\overline{N}_x} \sum_m \exp(\frac{2\pi i}{\overline{N}_x} m (k' - k)) \overline{U}(k, k') \\
    &= \tilde{\delta}_{k, k'} \overline{U}(k, k').
\end{eq}
Here, $\hat{\delta}$ is the $\delta$-function modulo $\overline{N}_x = N_x / M_x$, which equals one for $k' = k + l \overline{N}_x$, integer $l$, and vanishes otherwise.

An initial plane wave with momentum $p$ becomes after $\Delta t$ a superposition of momenta $p_l = p + 2\pi l / \Delta x$. The wave function in momentum space $\psi(\Delta t, q)$ vanishes for all momenta $q$ different from $p_l$. This continues after an arbitrary number of $\Delta t$ steps. Correspondingly, in the Heisenberg picture, one has for the momentum operator in Fourier space
\begin{eq}
 P_H(\Delta t)(k, k') = \sum_{k''} \overline{U}^{-1} (k, k'') \frac{2\pi k''}{N_x \epsilon} \overline{U}(k'', k').
\end{eq}
This operator has non-zero elements only for $k' = k + l \overline{N}_x$. One concludes that momentum is conserved $\mod 2\pi/\Delta x$. In sect. \labelcref{sect:densityMatrixAndCoarseGraining} we will explicitly construct a coarse grained momentum operator which commutes with the Hamiltonian.

For large enough $\Delta t$ and $n_{tot}$, it may happen that for the stochastic automaton $\overline{U}$ becomes approximately invariant under translations by $\epsilon$. The breaking of translation  symmetry by the distribution of scattering points may average out. In this case we can repeat the steps above for $\Delta x = \epsilon$. Momentum becomes a conserved quantity in this case. This would bring the RPCA even closer to the Dirac fermion.

\subsection{Time evolution of momentum distribution}

\begin{figure}[t]\centering
    \includegraphics[width=8.5cm]{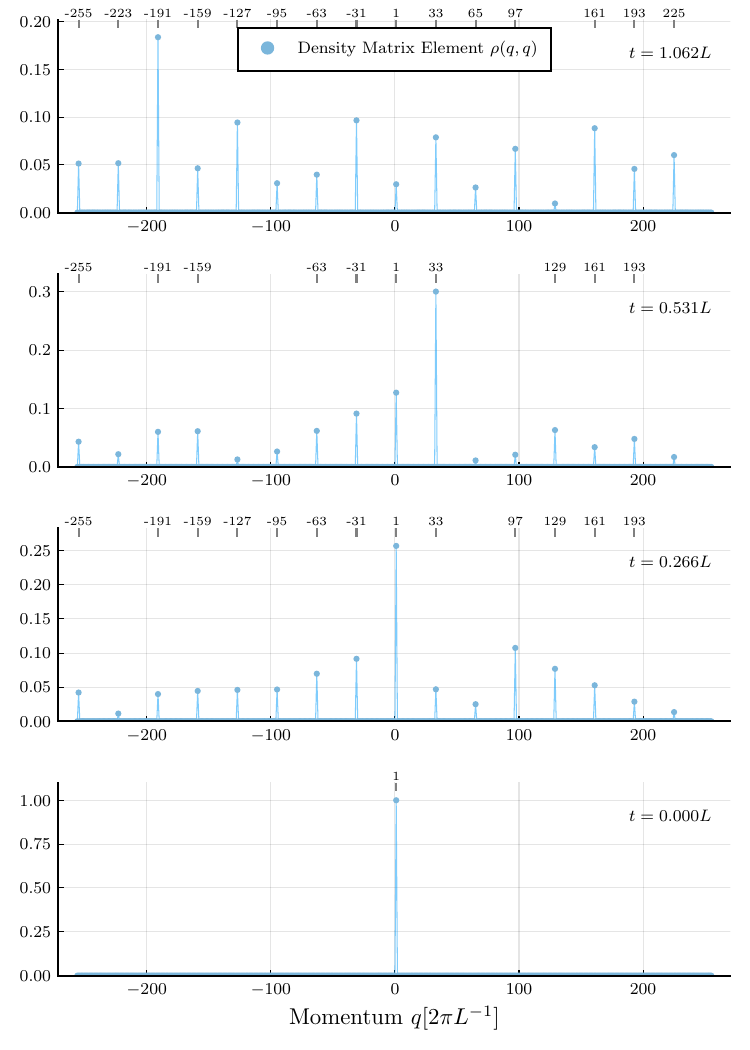}
    \caption{Momentum distribution at four different time steps, evolved with a periodic random PCA, model B \labelcref{eq:pRPCA_params}. We indicate the probabilities for given momenta, as encoded in the diagonal elements $\rho(q,q)$ of the density matrix in the momentum basis. The initial plane wave with a sharp momentum distribution evolves into a superposition of momentum states with momenta $(\overline{N}_x + 1)(2\pi / L)$, $m$ integer, $\overline{N}_x = N_x / M_x = 32$. We indicate the momentum values of the prominent momentum peaks at the top of each subfigure.}
    \label{fig:periodicModel_momentum_timeEvolution}
\end{figure}

We solve numerically the discrete evolution equation \labelcref{eq:TwoStepTimeEvolution} for the random probabilistic cellular automaton, using a given fixed random distribution of scattering points. In fig. \labelcref{fig:periodicModel_momentum_timeEvolution} we display the momentum distribution for a periodic RPCA (model B), with parameters
\begin{eq} \label{eq:pRPCA_params}
    \overline{N}_x = \frac{L}{\Delta x} = 32, \quad &\overline{N}_t = \frac{T}{\Delta t} = 128,\\ \quad n_{tot} = 16, \quad &\overline{n} = \frac{1}{17},\\
    M_x = \frac{\Delta x}{\epsilon} = 16, \quad  &M_t = \frac{\Delta t}{\epsilon} = 17.\\
\end{eq}
The distribution of the $16$ scattering points is shown in fig. \labelcref{fig:ism_example_trajectories}. We start at $t=0$ with an initial wave function given by a momentum eigenstate $\psi(0, x) = \psi_p(0, x), p = 2\pi / L$. In momentum space this sharp momentum state is a $\delta$- distribution. At later times we see how contributions with different momenta are generated during the evolution.
All these momenta correspond to the same coarse grained momentum, being equal $\mod 2\pi/\Delta x$.
The numerical solution clearly reproduces the conservation of the coarse grained momentum.

This conservation law for RPCAs may not easily be visible without the quantum formalism. Conserved quantities provide an obstruction for an approach to a homogeneous equilibrium state. Such a state could only be reached by choosing the initial wave function as an eigenstate of the coarse grained momentum with eigenvalue zero.
Conserved quantities are a robust way for storing memory of initial conditions even for rather complex automata and initial probability distributions.

\section{Energy observable} \label{sect:energy}

For quantum systems the energy is a central observable. This extends to probabilistic cellular automata, for which this observable may be less familiar. The operator for the energy is given by the Hamiltonian $H$, as defined by the meso-step evolution operator $\overline{U}$ in eq. \labelcref{eq:hermitianHamiltonian} with $U(t)$ given by \labelcref{eq:TwoStepTimeEvolution}. The real eigenvalues of the hermitian operator $H$ are the possible measurement values of the energy. In our discrete setting, the spectrum of eigenvalues of $H$ is discrete. We recall that $H$ does not depend on time. By its definition, $H$ commutes with $\overline{U}$. In the Heisenberg picture, one has for integer $n_t$
\begin{eq}
 H_H(n_t \Delta t) = H.
\end{eq}
The energy is therefore a conserved quantity.

\subsection{Periodic evolution for stationary states}

The description of probabilistic cellular automata as quantum systems leads to a striking feature. If we start at initial time $t=0$ with an eigenstate of $H$,
\begin{eq}\label{eq:energyEigenstate}
 H \psi_n(0, x) = E_n \psi_n(0, x),
\end{eq}
the time evolution for $t=n_t \Delta t$ is very simple
\begin{eq}
 \psi_n(t, x) = \exp(-i E_n t) \psi_n(0, x).
\end{eq}
Up to an overall phase, the distribution in space is the same for all $t=n_t \Delta t$. This phase drops out for the probability distribution or the mean occupation number of right- or left-movers
\begin{eq} \label{eq:equalMeanOccupationNumbers}
 \langle n_{R,L}(n_t \Delta t, x) \rangle = \langle n_{R,L}(0, x) \rangle.
\end{eq}

The phase remains visible in the separate real and imaginary parts of the complex wave function, and therefore in the probabilities $w_\gamma(t,x)$ for finding at $x$ a particle of type $\gamma$, or in the corresponding mean occupation number for this particle $\langle n_\gamma (t, x) \rangle = w_\gamma(t, x)$. The initial mean occupation numbers reappear after a full period $\langle n_\gamma(t_{in} + 2\pi / E_n, x) \rangle = \langle n_\gamma (t_{in}, x) \rangle$.

For our random automaton, the mean occupation numbers deviate substantially from the initial values after a certain number of steps of size $\epsilon$. Nevertheless, there exist particular initial distributions for $\left<n_R(x)\right>$ and $\left<n_L(x)\right>$ which reappear precisely after a mesoscopic time step $\Delta t$. In fig. \labelcref{fig:ism_qm_red_occupationNum_timeEvolution} we follow the time evolution of $\left< n_{R1}(x) - n_{L1}(x) \right>$ for one of the energy eigenstates discussed later in this section.  One observes a change of the distribution of red right- and left-movers after the first updating steps. After $\Delta t/ \epsilon=17$ time steps the original distribution reappears, now shifted in space. We display exemplarily $t= 544 \epsilon = 32 \Delta t $. After a full period, for $t= 1632 \epsilon = 96 \Delta t $, the initial distribution is recovered.

This generic periodic behavior in $\Delta t$ for particular initial probability distributions would be rather hard to guess without the quantum formulation at hand.
This is a simple, striking example for the usefulness of the quantum formalism for cellular automata.
We emphasize that this phenomenon can be observed by updating the probability distribution in the real formulation. The use of wave functions is convenient in order to understand what happens, but not mandatory for the presence of this periodic behavior of suitable probability distributions.

\subsection{Energy spectrum and eigenstates}

For the Dirac fermion, $H$ commutes with $P$ and we can find simultaneous eigenfunctions to $H$ and $P$. Since the hermitian $2N_x \times 2N_x$ matrix $H$ has $2N_x$ eigenvalues and $P$ has $N_x$ different eigenvalues $p$, we expect two energy eigenvalues for each $p$. Without subtraction of the constant part $m$, time reversal symmetry implies that both $E(p)$ and $-E(p)$ belong to the energy spectrum. Parity implies the same energy for $p$ and $-p$. For the random automaton these issues are more complex, since the explicit form of $\overline{U}$ or $H$ is not known.

The question arises how to find for the random automaton the wave functions and associated probability distributions which correspond to energy eigenstates. We are also interested in the energy spectrum of the random automaton, which may be compared to the one for the discrete quantum mechanics for the Dirac fermion. The Hamiltonian is a complex $2N_x \times 2N_x$ matrix with up to $2N_x$ different eigenvalues. For large $N_x$ direct diagonalization of $H$ becomes difficult. For periodic RPCAs we may exploit the fact that $H$ exhibits periodicity in position space
\begin{eq}
 H(x, x') = H(x + \Delta x, x' + \Delta x).
\end{eq}
This leads to a block diagonal structure of $H$ in momentum space that we use for diagonalization below.
We may in addition realize parity conservation and time reversal invariance by imposing additional constraints on the distribution of scattering points in the region $(\Delta t, \Delta x)$. For large enough $\Delta t$, it is also possible that these discrete symmetries are realized approximately.
Finally, if the number $n_{tot}$ of scattering points is not too large we can construct explicitly particular energy eigenstates as single-orbit states, see below.

\subsection{Transition element}

\begin{figure}[t]\centering
    \includegraphics[width=8.5cm]{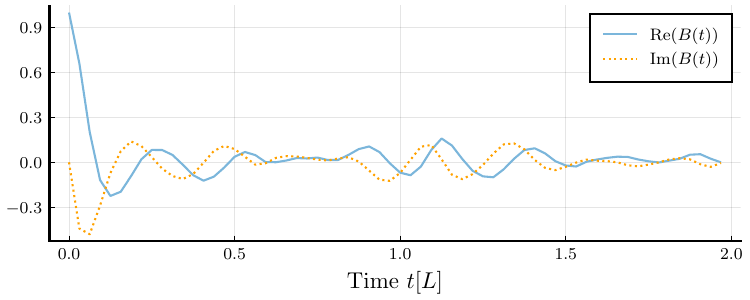}
    \caption{Transition element $B(t)$ of the Brownian PCA, model A \labelcref{eq:brownianModelParams}, with an initial plane wave. One observes a type of damped oscillations, rather than a smooth decay towards some equilibrium state.}
    \label{fig:brownianModel_autocorrelation}
\end{figure}

\begin{figure}[t]\centering
    \includegraphics[width=8.5cm]{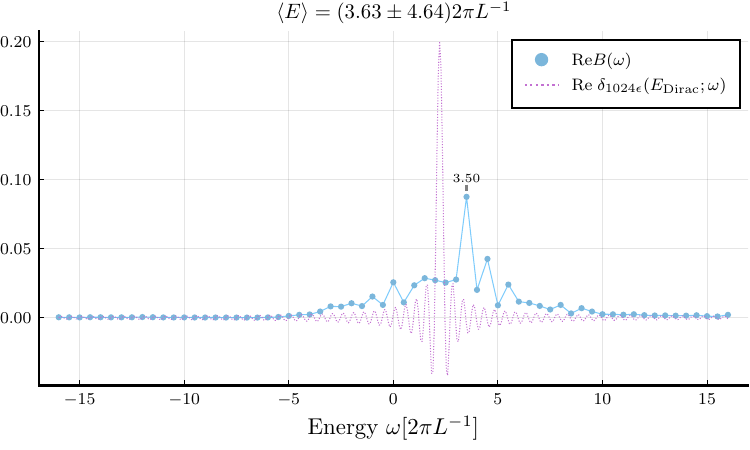}
    \caption{Short term energy spectrum $Re B(\omega)$ of the Brownian PCA, model A \labelcref{eq:brownianModelParams}, with an initial plane wave. For $\overline{N}_t = 1024$ intervals $\Delta t$ we compare this with the discrete generalization of the $\delta-$ function, which characterizes a single eigenstate. For the Brownian automaton the initial plane wave is a superposition of different energy eigenstates. The peak of the energy distribution is close to the mean value $\langle E \rangle$.}
    \label{fig:brownianModel_shortTermEnergySpectrum}
\end{figure}

A first approach for finding the spectrum of $H$ employs a Fourier type transform to frequency space for the transition element
\begin{eq}
 B(t;\overline{t}) = \sum_x \psi^\dagger (\overline{t}, x) \psi(\overline{t} + t, x).
\end{eq}
We plot the transition element $B(t) = B(t; 0)$ for the Brownian automaton (model A) in fig. \labelcref{fig:brownianModel_autocorrelation}. The beginning oscillating behavior is damped. We perform a Fourier transform to frequency space in order to extract the energy distribution. For this purpose we select $t$ and $\overline{t}$ as integer multiples of $\Delta t$, with $t$ extending over $\overline{N}_t + 1$ discrete values. We define the discrete Fourier transform to frequency space by
\begin{eq}
    B(\omega) = \frac{1}{\overline{N}_t + 1} \sum_{t} e^{i \omega t} B(t; \overline{t}).
\end{eq}
The real part of $B(\omega)$ for the Brownian automaton (model A) is shown in fig. \labelcref{fig:brownianModel_shortTermEnergySpectrum} for $\overline{t}=0$. One finds a broad peak at small frequencies, with almost no contribution of frequencies $|\omega| \gtrsim 10 \cdot (2\pi / L)$.

For an extraction of information on the energy spectrum we expand $\psi(\overline{t}, x)$ in energy eigenstates \labelcref{eq:energyEigenstate},
\begin{eq}
 \psi(\overline{t}, x) = \sum_n \alpha_n \psi_n(x).
\end{eq}
For an orthonormal system of eigenfunctions,
\begin{eq}
 \sum_x \psi^\dagger_m(x)\psi_n(x) = \delta_{mn},
\end{eq}
one finds
\begin{eq}
 B(t; \overline{t}) = \sum_n |\alpha_n|^2 \exp(-i E_n t).
\end{eq}
This yields in frequency space
\begin{eq}
 B(\omega) &= \sum_n |\alpha_n|^2 \delta^{\overline{N}_t}(\omega, E_n),
\end{eq}
with
\begin{eq}
 \delta^{\overline{N}_t} (\omega, E_n) &= \frac{1}{\overline{N}_t + 1} \sum_{t} \exp(i (\omega - E_n) t).
\end{eq}
The frequencies $\omega$ are periodic with $\omega$ and $\omega + 2\pi / \Delta t$ identified. We consider $\overline{N}_t + 1$ discrete values $\omega = 2\pi k_\omega / ((\overline{N}_t + 1) \Delta t)$, with integers $k_\omega$ in the range $\left[ -\frac{\overline{N}_t}{2}, \frac{\overline{N}_t}{2}\right]$. In general, $B(\omega)$ depends on $\overline{t}$ and $\overline{N}_t$.

The coefficients $|\alpha_n|^2$ do not change during the evolution and are therefore independent of $\overline{t}$, due to
\begin{eq}
    \psi(\overline{t} + t, x) &= \sum_n \alpha_n e^{-i E_n t} \psi_n (x) = \sum_n \alpha'_n \psi_n(x),\\
    |\alpha'_n|^2 &= |\alpha_n|^2.
\end{eq}
They reflect the probabilities for the different energies $E_n$ of the initial state. This energy distribution is preserved in time.
The conserved mean energy and energy fluctuation are given by
\begin{eq}
    \langle E \rangle &= \sum_n |\alpha_n|^2 E_n,\\ \langle E^2 \rangle - \langle E \rangle ^2 &= \sum_n |\alpha_n|^2 E_n^2 - \langle E \rangle^2.
\end{eq}

The extraction of the spectrum depends, however, on $\overline{t}$ and $\overline{N}_t$ through the range of the  summation for $\delta^{\overline{N}_t}$.
If the range of $t$ is symmetric around zero (i.e. $\overline{t}$ in the middle of the interval covered by $t$) the imaginary part of $\delta^{\overline{N}_t}$ vanishes.
For large $\overline{N}_t$ the function $\delta^{\overline{N}_t}(\omega, E_n)$ decreases rapidly for $|\omega - E_n| > \Delta t^{-1}$ from the maximal value $1$, that it takes for $\omega = E_n$. On the other hand, it remains close to one for $|\omega - E_n| \ll (\overline{N}_t \Delta t)^{-1}$. We conclude, that $B(\omega)$ yields a smeared energy distribution. It cannot resolve energy differences smaller than $(\overline{N}_t\Delta t)^{-1}$.

In fig. \labelcref{fig:brownianModel_shortTermEnergySpectrum} we indicate $\Re (\delta^{\overline{N}_t} (\omega, \overline{E}))$ for the energy as predicted for a Dirac fermion $\overline{E}  = \sqrt{m^2 + p^2} - m \approx 2,22 \cdot (2\pi / L)$ and $\overline{N}_t = 64, \overline{t}=0$. We conclude that $\Re B(\omega)$ is substantially broader than $\delta^{\overline{N}_t} (\omega, \overline{E})$. The initial plane wave state therefore involves an extended range of energy eigenvalues of the Brownian automaton.


\subsection{Energy variance and variational approach to energy eigenstates}

For a given initial state we can employ the evolution for four mesoscopic time steps $\Delta t$ in order to analyze how close it is to an energy eigenstate. One computes the variance (mean quadratic fluctuation) of a simple function of the Hamiltonian.

Let us define the operator
\begin{eq}
 \tilde{H} = \frac{1}{\Delta t} \sin(H \Delta t).
\end{eq}
Its expectation value obeys
\begin{eq} \label{eq:H_tilde_expectationValue}
\langle \tilde{H} \rangle(t) &= \sum_x \psi^{\dagger}(t, x) \frac{1}{\Delta t} \sin(\Delta t H) \psi(t, x) \\
&= \sum_x \psi^{\dagger}(t, x) \frac{i}{2\Delta t} \left(e^{-i\Delta t H} - e^{i\Delta t H}\right) \psi(t, x) \\
&= \sum_x \psi^{\dagger}(t, x) \frac{i}{2\Delta t} \left(\psi(t + \Delta t, x) - \psi(t - \Delta t, x)\right).
\end{eq}
In the last line we recognize a discrete time derivative. Since $[\tilde{H}, H] = 0$, the expectation value $\langle \tilde{H} \rangle$ actually does not depend on time.

Similarly, we may compute
\begin{eq} \label{eq:H_tilde_variance}
 \langle \tilde{H}(t)^2 \rangle = -\frac{1}{4\Delta t^2}\sum_x &\psi^\dagger(t, x) [\psi(t + 2\Delta t, x) \\
  &- 2\psi(t, x) + \psi(t - 2\Delta t, x) ]
\end{eq}
We define the variance of $\tilde{H}$,
\begin{eq} 
 D = \langle \tilde{H}^2 \rangle - \langle \tilde{H} \rangle^2 = \langle (\tilde{H} - \langle \tilde{H} \rangle)^2 \rangle.
\end{eq}
For eigenfunctions of $\tilde{H}, \tilde{H}\psi_n = \tilde{E}_n \psi_n$, one has $D=0$, and vice versa $D=0$ implies that $\psi$ is an eigenfunction of $\tilde{H}$. Eigenfunctions of $\tilde{H}$ are also eigenfunctions of $H$ with eigenvalues related by $\tilde{E}_n = \frac{1}{\Delta t} sin(E_n \Delta t)$.

If one finds a state with $D=0$, one has established an eigenstate of the Hamiltonian. Correspondingly, the size of $D$ measures how far a given initial state is from an eigenstate of $H$. As long as energies $E_n$ with $|\Delta t E_n| \ll 1$ dominate one can take $D$  as a direct measure for the variance of $H$, $D = \langle H^2 \rangle - \langle H \rangle ^2 - \frac{\Delta t^2}{3}(\langle H^4 \rangle - \langle H \rangle \langle H^3 \rangle) + ...$
Similarly, we can approximately determine the expectation value of $H$, $\langle \tilde{H} \rangle = \langle H \rangle - \langle H^3 \rangle \Delta t^2 / 6 + ...$

One can use the values of the wave function for four steps in $\Delta t$ for a variational approach to find energy eigenvalues, such as machine learning techniques. To this end, one might choose some trial wave function and calculate $D_\alpha$ by evolving the automaton from $t=0$ to $t=4\Delta t$ and calculate $D_\alpha$, taking in eqs. \labelcref{eq:H_tilde_expectationValue}, \labelcref{eq:H_tilde_variance} $t=2\Delta t$. Optimization of $D_\alpha$ may then yield approximate eigenfunctions of $H$. We have not taken this path. Instead we calculate eigenfunctions using a numerical diagonalization of the step evolution operator, which is feasible for sufficiently small systems. In this case, computation of $D=0$ can serve as a verification that a proposed state actually is an eigenfunction of the Hamiltonian.

\subsection{Static states}

Probabilistic cellular automata with a time-independent deterministic updating rule typically admit many static states. For static states the probability distribution and wave function do not change with time. In our context this applies to the mesoscopic level, such that static wave functions obey $q_\gamma(t + n_t \Delta t, x) = q_\gamma(t, x)$. A general construction rule allows us to classify the static states.

A finite, invertible cellular automaton with a time independent updating is a clock system, and the PCA therefore a probabilistic clock system \cite{CWPW2020}.
A clock system is characterized by its orbits or clocks.
Let us start at $t_{in}$ with a sharp state for which a particle of type $\gamma_1$ is located at a given position $x_1$.
According to the updating rule it will be found at $t_{in} + \epsilon$ at position $x'_1$ and have color $\gamma'_1$, and so on for further steps.
After a number $N_1$ of time steps it will return to the position $x_1$ with color $\gamma_1$. The ensemble of one-particle configurations $(x_1, \gamma_1), (x'_1,  \gamma'_1), (x''_1, \gamma''_1)...$ constitutes the orbit associated to $(x_1, \gamma_1)$.
The length of the orbit is given by the number of one-particle configurations in the ensemble or orbit, and equals $N_1$.
The maximal length of the orbit amounts to the total number of configurations of the automaton.
In our case it equals $N_{max} = 4N_x$. For $N_1 < N_{max}$ we can start with a new configuration $(x_2, \gamma_2)$, which does not belong to the orbit of $(x_1, \gamma_1)$. Following the same procedure we construct the orbit of $(x_2, \gamma_2)$ with length $N_2$ obeying $N_2 \leq N_{max} - N_1$.
Repeating the procedure decomposes the total number of one-particle configurations into $m$ orbits with length $N_m, \sum_m N_m = N_{max}$. The evolution within the members of a given orbit proceeds independently of the other orbits.

For periodic PCAs we may define a reduced orbit which ends if a one-particle trajectory starting at $(x_m, \gamma_m)$ has reached the configuration $(x_m + s\Delta x, \gamma_m)$ for some integer $s$. For $s=0$ and $s=\overline{N}_x$ the reduced orbit coincides with the full orbit. For $s \neq 0$ periodicity implies that the orbit continues in the same way, now shifted by $s \Delta x$. For $0 < s < \overline{N}_x$ the full orbit can be constructed by attaching reduced orbits to each other until $\nu_m s \Delta x = \mu_m L$, with $\nu_m$ and $\mu_m$ integers. The length of the reduced orbits counts the number $n_s$ of time steps needed to reach $(x_m + s\Delta x, \gamma_m)$. The length of the full orbit is given by $N_m = n_s \nu_m$.

The configurations belonging to the orbit $m$ may be denoted collectively by $(x^{(m)}, \gamma^{(m)})$, i.e. the members of this set are the pairs $(x_m,\gamma_m)$, $(x'_m,\gamma'_m)$ etc...
Static wave functions, and correspondingly static probability distributions, obtain by associating at $t_{in}$ the same value to all components which correspond to members of a given orbit
\begin{eq} \label{eq:constructionStaticWaveFunction}
    q_{\gamma^{(m)}} (t_{in}, x^{(m)}) = q^{(m)}.
\end{eq}
Here we can identify $q^{(m)}$ with the component of the initial wave function $q_{\gamma_{m}} (t_{in}, x_{m})$ for the state $(x_m, \gamma_m)$ from which we have constructed the orbit $m$.

In order to show that a wave function with initial value \labelcref{eq:constructionStaticWaveFunction} is static, we note that the component of the wave function $q_{\gamma^{(m)}} (t_{in} + \epsilon, x^{(m)})$ for any given $(x^{(m)}, \gamma^{(m)})$ belonging to the orbit $m$ equals the component of the initial wave function for the configuration from which the configuration $(x^{(m)}, \gamma^{(m)})$ has originated.
Since this original state belongs to the orbit $m$, it is given by $q^{(m)}$. By virtue of eq. \labelcref{eq:constructionStaticWaveFunction} one infers time-independence.
This holds similarly for all components associated to the orbit $m$. Thus the wave function remains invariant under the updating.

All normalized linear superpositions of static wave functions are again static wave functions.
The static wave functions form a subspace of $\mathbb{R}^m$ with coordinates $q^{(m)}$ and the additional normalization condition
\begin{eq}
    \sum_m N_m (q^{(m)})^2 = 1.
\end{eq}
The dimension of $\mathbb{R}^m$ equals the number of independent orbits.
There typically exists a large number of static wave functions.

\begin{figure}[t]\centering
    \includegraphics[width=8.5cm]{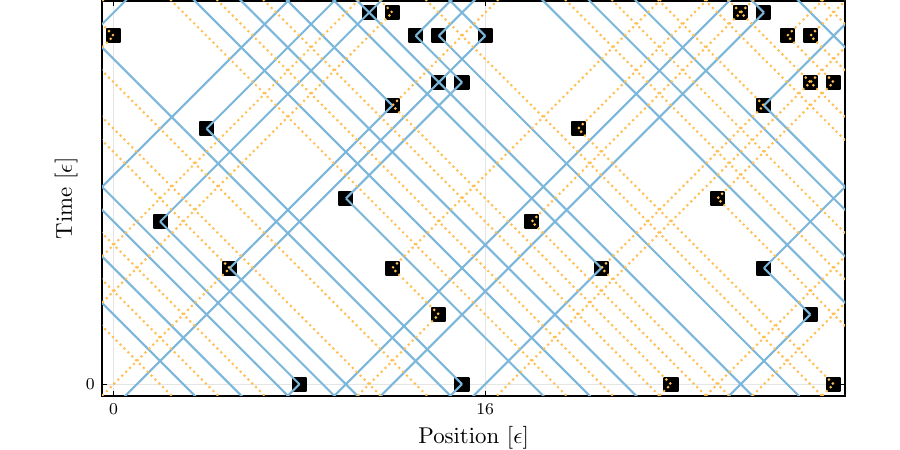}
    \caption{Orbits for the periodic cellular automaton, model B. The blue and orange orbit can be followed by periodic continuation in $\Delta x$ and $\Delta t$. The two orbits shown are shifted by $\Delta x$ to each other.}
    \label{fig:ism_orbit_structure}
\end{figure}

We show a typical reduced orbit in fig. \labelcref{fig:ism_orbit_structure}. The scattering points are indicated by the black squares, which are continued periodically in $x$ with $\Delta x = 16 \epsilon$, and in $t$ with $\Delta t = 17 \epsilon$. The particular distribution of scattering points for a periodic RPCA is the same as for fig. \labelcref{fig:ism_example_trajectories} and corresponds to model B.
The trajectory is indicated by straight lines that change direction at the scattering points. The lines are continued periodically in $\Delta x$ and $\Delta t$. The members of the orbit are at points $x$ which are hit by the periodic trajectory at $t=0,17,34,...,$ with colors $\gamma_m$ not indicated in the figure. The reduced orbit closes after $2 \Delta x, s = 2$. The length of this reduced orbit is $n_s = 18$. After the time $n_s \Delta t$ the trajectory repeats its pattern, by that time however shifted by $2 \Delta x$. A trajectory which starts at a position shifted by $2\Delta x$ is hence part of the same \textquote{full orbit}. For orbits with $s=2$ there exists a similar orbit of equal length obtained by a shift by $\Delta x$.

\subsection{Construction of energy eigenstates}

One can find explicit recipes for the construction of energy eigenstates for a small enough number of one-particle configurations and scattering points. This can be used to find explicitly all stationary states and to determine the complete energy spectrum.
As the number of possible configurations increases to large numbers, the practical use of this approach may find its limitations.
Nevertheless, this construction principle gives insight into the general structure of the energy spectrum of cellular automata.
The construction principle is based on single-orbit states.

The updating rule does not mix different (full) orbits. This means that the Hamiltonian is block-diagonal in a basis organized by the different orbits. The components $q_{\gamma^m}(t, x^{(m)})$ evolve independently of the components associated to different orbits. After $N_m$ steps $\Delta t$ the configuration returns to its original value.
Every component corresponding to $(\gamma, x)$ belonging to the orbit $m$ obeys
\begin{eq}
    q_{\gamma^{(m)}} (t + N_m \Delta t, x^{(m)}) = q_{\gamma^{(m)}} (t, x^{(m)}).
\end{eq}
We can therefore construct wave functions with a periodic time evolution as \textquote{single-orbit states} by setting $q_{\gamma^{(n)}}(t_{in}, x^{(n)}) = 0$ for $n \neq m$.
The single-orbit states show a periodic time evolution with period $N_m \Delta t$.
The linearly independent single-orbit states constitute a complete basis for the general real wave function.
For a given orbit $m$ one has $N_m$ linearly independent periodic single-orbit wave functions.
They can be constructed as linear combinations of the sharp wave functions for the $N_m$ one-particle configurations $(x^{(m)}, \gamma^{(m)})$ belonging to the orbit.
In turn, the number of single-orbit wave functions for all orbits is given by $\sum_m N_m = N_{max}$ and equals the number of basis functions for the general wave functions.

For our particular updating rule and the simple complex structure $\psi_R = q_1 + i q_2, \psi_L = q_3 + i q_4$, the notion of orbits and single-orbit wave functions carries over to the complex formulation.
For the starting point $(x_m, \gamma_m)$ of the orbit $m$ we define the associated starting point $(x_m, \overline{\gamma}_m)$, where the pair $(\gamma_m, \overline{\gamma}_m)$ is given by $(1, 2), (2, 1), (3, 4), (4, 3)$.
After $N_m$ steps both $(x_m, \gamma_m)$ and $(x_m, \overline{\gamma}_m)$ have returned to the same values. If $\gamma_m$ and $\overline{\gamma}_m$ correspond both to right-movers (the pairs $(1, 2), (2, 1)$), they encounter the same scattering points, such that $(x_m, \overline{\gamma}_m)$ is mapped to $(x_m, \overline{\gamma}'_m)$. Furthermore $\overline{\gamma}'_m$ corresponds again to a right-mover. Since invertibility forbids $\overline{\gamma}'_m = \gamma_m$, the only possibility is $\overline{\gamma}'_m = \overline{\gamma}_m$. The same holds for initial left-movers. The automaton property then implies for the complex one-particle wave function $\psi^{(m)}$ the periodicity property
\begin{eq} \label{eq:periodicityProperty}
    \psi_\alpha^{(m)} (t+N_m\Delta t, x) &= exp\{-i N_m \Delta t H\} \psi_\alpha^{(m)} (t, x)\\
    &= \psi_\alpha^{(m)} (t, x).
\end{eq}

This periodicity property determines the energy spectrum.
Since the evolution does not mix single-orbit states with different $m$, the Hamiltonian is block diagonal in a basis of single-orbit states. For a given orbit $m$ the block $H^{(m)}$ is a Hermitian $N_m \times N_m$-matrix with $N_m$ real eigenvalues $E^{(m)}_{k}$ obeying the condition
\begin{eq} \label{eq:singleOrbitEnergies}
    E^{(m)}_{k} = \frac{2\pi k}{N_m \Delta t}\ ,\quad |k| \leq \frac{N_m}{2},
\end{eq}
with integer $k$. The restriction $|k| \leq N_m/2$ reflects the periodicity in $E$ of $2\pi / \Delta t$.
We conclude that the block for single-orbit states for a given orbit $m$ has an equidistant energy spectrum.
The number of energy levels equals $N_m$.
Combination with energy eigenvalues of the single-orbit  states for different $m$ can result in a rather rich energy spectrum if $N_{max}$ is very large.
For the rather moderate values of $N_{max}$ used for our numerical simulations we expect that substantial restrictions on the energy spectrum remain.
The energy spectrum obtained by combining the energy eigenvalues of all single-orbit states is complete, corresponding to the completeness of the \textquote{single-orbit basis}.
Indeed, the number of energy levels (which may be degenerate) amounts to $\sum_m N^{(c)}_m = N^{(c)}_{max}$, which is the dimension of the (finite) Hilbert space. In the complex picture two ``real orbits'' are combined into one ``complex orbit'', such that $N^{(c)}_{max}=N_{max}/2=2N_x$. (We often omit the superscript $(c)$.)

Eigenstates to the energy eigenvalues $E^{(m)}_k$ can be constructed in a straight-forward way.
For this purpose we order the one-particle configurations (corresponding to sharp states) $\tau = (x, \gamma)$. For a given orbit $m$ we denote by $\tau_0^{(m)}$ the configuration $(x_m, \gamma_m)$ used to start the orbit.
The configuration obtained by $j$ updatings is denoted by $\tau_j^{(m)}$, with $\tau_{N_m}^{(m)} = \tau_0^{(m)}$. Then the eigenstate for $E^{(m)}_k$ has at $t_{in}$ the nonzero components
\begin{eq}
    \psi_{\tau_j^{(m)}} (t_{in}) = \varphi_0 exp\{\frac{2\pi i k j}{N_m}\}\ ,\ |\phi_0|^2 = \frac{1}{N_m}.
\end{eq}
This implies indeed
\begin{eq}
    \psi_{\tau_j^{(m)}} (t + \Delta t) = \psi_{\tau_{j-1}^{(m)}} (t)= \exp\{-i E_k^{(m)} \Delta t\} \psi_{\tau_j^{(m)}} (t),
\end{eq}
as appropriate for the eigenstate.

Single-orbit wave functions for orbits with $N_m \ll N_{max}$ are rather sparse since $\psi_\tau$ vanishes for all $\tau$ except the ones belonging to the orbit.
They will be far from the smooth wave functions needed for a possible continuum limit. For orbits with large $N_m$ and $|k|$ of the order $N_m$ one expects strong variations of the wave function on distance scales $\lesssim \Delta t$. The best chances for smooth eigenfunctions of $H$ correspond to $N_m$ near $N_{max}$ and small $k$. The automaton for a free massless Dirac fermion and $\Delta t = \epsilon$ realizes this setting, with $N_m = N_x = N^{(c)}_{max} / 2$ and energy equal to momentum $E(k) = p(k)$.

\subsection{Velocity and momentum for single-orbit states}

One can associate a velocity $v(m)$ to a given orbit $m$. This extends naturally to a velocity of single-orbit states. Single-orbit states may sometimes also be eigenstates of momentum or coarse-grained momentum. In this case one obtains a relation between velocity and momentum, and a dispersion relation for the relation between energy and momentum. For single-orbit states which are simultaneously eigenstates of energy and momentum this dispersion relation is linear.

Let us consider periodic automata for which the step evolution operator is invariant under space translations by $\Delta x$. For periodic boundary conditions the maximal value of $\Delta x$ equals $L = N_x \epsilon$, while our setting also includes the case $L=\overline{N}_x \Delta x$ with an integer number of $\Delta x$-intervals $\overline{N}_x$.
Once a one-particle configuration $\tau=(x, \gamma)$ is mapped by the evolution after a certain number $n_s$ of time steps $\Delta t$ to a point $(x+s \Delta x, \gamma)$, translation invariance in space and time implies that the trajectory has to repeat itself in the following, shifted in $x$ by $s \Delta x$. For $s$ larger than $\overline{N}_x$, the trajectory winds around the circle. We can now associate $n_s$ with the length of a reduced orbit.
This property allows us to associate to each sharp one-particle state an average velocity $v$.
Counting the number $n_s$ of $\Delta t$-steps needed for reaching $(n + s \Delta x, \gamma)$, the average velocity of the sharp single particle is given by
\begin{eq}
    v = \frac{s \Delta x}{n_s \Delta t}.
\end{eq}
It determines how fast a sharp particle propagates in time in the average. (For a single interval, $\Delta x = L$, the \textquote{stride} $s$ coincides with the winding number. Note, that $s \Delta x$ is not periodic in $L$).
For $s=0$ (no winding) the one-particle configuration is static in the average, $v=0$. For $s \neq 0$ the number $n_s$ is proportional to $s$ such that it plays no role at which $s$ the velocity is measured.
For a large number $\overline{N}_x$ of $\Delta x$-intervals and small $s$ the velocity becomes a local property.
For our automaton the maximal velocity equals $\pm 1$, defining the \textquote{light cone}. The average velocity of a sharp single-particle configuration does not depend on time.
One may start at the configuration $(x(\Delta t), \gamma(\Delta t))$ reached after a time step $\Delta t$. After $n_s$ steps $\Delta t$ this one-particle configuration has reached $(x(\Delta t) + s \Delta x, \gamma(\Delta t))$.

By virtue of time translation invariance by $\Delta t$ all points of the orbit constructed from $(x_m, \gamma_m)$ must have the same $v$.
The average velocity $v(m)$ is therefore a property of the orbit $m$.
We can therefore associate an average velocity to an arbitrary single-orbit wave function. This holds, in particular, for the energy eigenstates. The average for the velocity is taken over a time $n_s \Delta t$. We may define for a sharp one-particle state a generalized average velocity $v_w(t, x)$ by
\begin{eq}
    v_w(t, x^{(m)}) = \frac{\overline{x}^{(m)}(t + w \Delta t) - x^{(m)}(t)}{w \Delta t}.
\end{eq}
Here $\overline{x}_m (t + w \Delta t)$ adds to $x^{(m)} (t+wt)$ the number of windings times $L$.
The velocity $v_w$ differs for the different positions $x^{(m)}$ belonging to the orbit $m$, and it depends on time.
For all $x^{(m)}$ on the orbit one has, however
\begin{eq}
    v_{n_s} (t, x^{(m)}) = v(m).
\end{eq}

Consider next the possibility that single-orbit energy eigenstates are simultaneously momentum eigenstates. For an energy $E_k^{(m)}$ we want to find the allowed values of the momentum $p(E_k^{(m)})$. One finds a linear dispersion relation
\begin{eq} \label{eq:linearDispersionRelation}
    p(E_k^{(m)}) = (E_k^{(m)} - n_E \Delta E) / v(m).
\end{eq}
In order to establish this relation we use periodicity in space and time for a wave function that is simultaneously an eigenstate of $P$ and $H$,
\begin{eq}
    \psi_\alpha (t, x) = \psi_{0, \alpha} \exp\{i (px - E (t - t_0))\}.
\end{eq}
Such a wave function can be realized by a single-orbit state only if the orbit covers all positions $x$.
Orbits for which not all positions are reached are necessarily superpositions of momentum eigenstates.

Let us consider single-orbit states for which the wave function returns to itself when a reduced orbit is closed after $n_s$ time steps $\Delta t$,
\begin{eq} \label{eq:periodicSingleOrbitState}
    \psi_\alpha (t_0 + n_s \Delta t, x + s \Delta x) = \psi_{0, \alpha} \exp\{i p x\}.
\end{eq}
For $\Delta x < L$ this reflects a subset of energy eigenvalues. Nevertheless, our construction also covers the full orbit and arbitrary energies if one chooses $\Delta x = L, n_s = N_m$, where the integer $s$ is the winding number. Eq. \labelcref{eq:periodicSingleOrbitState} implies the relation
\begin{eq}
    p s \Delta x - E n_s \Delta t = -2\pi \overline{k},
\end{eq}
or
\begin{eq} \label{eq:linearDispersion_pc_E}
    p v = E_k^{(m)} - \frac{2\pi \overline{k}}{n_s \Delta t}.
\end{eq}
Here the integer $\overline{k}$ is chosen such that $E_k^{(m)}$ and $E_k^{(m)} - 2\pi \overline{k} / (n_s \Delta t)$ yield the same value of $p$, taking into account the periodicity of $p$. This establishes eq. \labelcref{eq:linearDispersionRelation} with
\begin{eq}
    \Delta E = \frac{2\pi v}{\Delta x}.
\end{eq}
Non-zero integers $n_E$ occur if the length of the orbit $n_s$ exceeds $N_x / \epsilon$.

If the step evolution operator is invariant by space-translations $\Delta x$, one can introduce a coarse grained momentum operator $\overline{P}$ which commutes with $H$. One can find simultaneous eigenstates of $H$ and $\overline{P}$, given for the eigenvalues $E$ and $\overline{p}$ by
\begin{eq}
    \psi_\alpha (t,x) = \sum_l \psi_{0, \alpha}^{(l)} \exp\{i (\overline{p} + \frac{2\pi l}{\Delta x})x - i E(t - t_0)\},
\end{eq}
with integer $l$ in the interval $[-\Delta x / (2\epsilon), \Delta x / (2\epsilon)]$.
Let us assume a single-orbit energy eigenstate which is simultaneously an eigenstate of the coarse grained momentum.
The dispersion relation is again linear, with a replacement in eq. \labelcref{eq:linearDispersionRelation} of $p$ by $\overline{p}$ and modified $\Delta E, n_E$. We have found numerically these relations for automata with suitable orbits.

\subsection{General eigenstates of energy and momentum}

This picture of simultaneous eigenstates of energy and coarse grained momentum is, however, not complete.
There is no need that a simultaneous eigenstate of $H$ and $\overline{P}$ is a single-orbit state. Neither is it guaranteed that every single-orbit energy eigenstate is an eigenstate of the coarse grained momentum. First, there may be different orbits with the same length $N_m$, and therefore the same spectrum of energy eigenvalues. The eigenstate of $\overline{P}$ can then be a linear combination of the single-orbit states.
This is what we have found typically for simulations of simple automata which have distinct orbits with the same length.
For full orbits with the same length and winding number the velocity $v$ is the same and the linear dispersion relation continues to hold.
A much richer structure arises if the same energy eigenvalue $E$ occurs in two orbits of different length or winding number.
In this case there is no unique velocity associated to this energy. The eigenstate of $\overline{P}$ can be a superposition of single-orbit states with different orbit-velocity. If this type of mixing of orbits with different $v$ is realized, one may find a non-linear dispersion relation. Such a non-linear relation may be expected, in particular, if the mixing of orbits depends on the energy eigenvalue $E$.

For systems close to a continuum limit the energy spectrum is almost continuous. Such systems have a very large number of states and may have many different very long orbits with different $v$. One expects many energy eigenvalues to occur for orbits of different length. Eigenstates of (coarse grained) momentum are typically no longer single-orbit states. There could then be simultaneous eigenstates of energy and momentum which are superpositions of single-orbit states for orbits with different length and no linear dispersion relation.

An energy eigenstate which is a superposition of two or more single-orbit states follows a periodic evolution, as given by the energy eigenvalue. This is a simple consequence of the superposition principle in quantum mechanics. This periodicity can no longer be understood in a simple way on the level of the time evolution of the probability distribution. Only for single-orbit states the periodicity finds a simple explanation on the level of probabilities. For linear combinations of single-orbit states the superposition law holds on the level of wave functions, but not for probability distributions. This demonstrates once more the important advantage of the quantum formalism with wave functions for the understanding of probabilistic cellular automata.

\subsection{Eigenstates of coarse grained momentum and energy}\label{sect:combinedEigenstates}

The eigenstates of the coarse grained momentum with eigenvalue $\overline{p}$ are plane waves -- they are the eigenstates of momentum for the values of $p = \overline{p} \operatorname{mod} 2\pi / \Delta x$. Since the coarse grained momentum is conserved, one can find simultaneous eigenstates of energy and coarse grained momentum. For this purpose one has to diagonalize the evolution operator restricted to the eigenstates of coarse grained momentum . If the dimension of this subspace is not too large, this diagonalization can be done explicitly, as demonstrated here by simple examples.

For the periodic stochastic automaton the mesoscopic evolution operator in momentum space takes a block diagonal form
\begin{eq}
    U(q, q') &= U(\overline{q}, Q; \overline{q}', Q') = W(\overline{q}; Q, Q') \delta_{\overline{q}, \overline{q}'}.
\end{eq}
This is a direct consequence of the conservation of the coarse grained momentum $\overline{q}$. The block matrices $W(\overline{q}; Q, Q')$ depend, in general, on $\overline{q}$ such that $U$ does not take a direct product form.
For a given $\overline{q}$ one may diagonalize the unitary matrix $W(\overline{q})$ by a basis transformation,
\begin{eq}
    W'(\overline{q}) = \mathrm{diag}(e^{-i \alpha_\lambda(\overline{q})}),
\end{eq}
with eigenvalues directly related to the eigenvalues of the Hamiltonian $E_\lambda (\overline{q})$,
\begin{eq}
    \alpha_\lambda (\overline{q}) = E_\lambda (\overline{q}) \Delta t, \quad E_n = E_\lambda (\overline{q}).
\end{eq}
The dimension of the matrix $W(\overline{q})$ depends on the number of sites in the interval $\Delta x$, i.e. $M_x = \Delta x / \epsilon$, and on the number of internal degrees of freedom in the complex formulation.

The $2M_x \times 2M_x$ matrix $W(\overline{q})$ can be extracted by starting at initial time $t=0$ with plane waves with momenta $\overline{q} + Q'$. One computes at $\Delta t$ the Fourier components of $\psi(\Delta t)$ with momenta $\overline{q} + Q$. For the plane waves at $t=0$ we use for each $\overline{q} + Q'$ two wave functions $\psi_\beta$ with internal components $\begin{pmatrix}1\\0\end{pmatrix}$ and $\begin{pmatrix}0\\1\end{pmatrix}$, respectively.
For a given initial $\overline{q} + Q'$ and $\beta$ one finds then 
\begin{eq}
    \psi_\alpha (\Delta t, \overline{q} + Q) = W_{\alpha\beta} (\overline{q}; Q, Q').
\end{eq}

\begin{figure}[t]\centering
    \includegraphics[width=8.5cm]{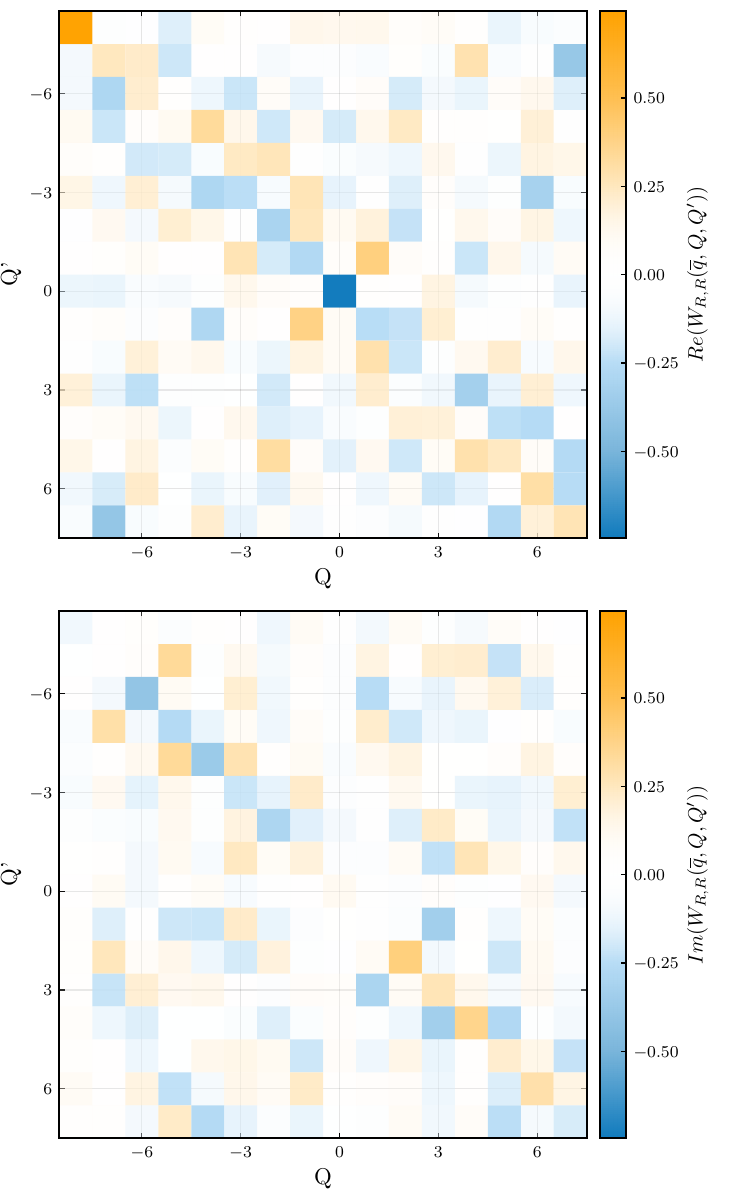}
    \caption{Reduced mesoscopic evolution operator of model B \labelcref{eq:pRPCA_params}. We display the submatrix $W_{R, R}(\overline{q}; Q, Q')$ for the right-moving component for $\overline{q} / (2\pi / L) = 1$. The dominant components are the real parts of the elements $(0,0)$ and $(-8, -8)$, which are opposite in sign.}
    \label{fig:U_p_reduced}
\end{figure}

We have performed the diagonalization for the parameters of model B \labelcref{eq:pRPCA_params}. The matrix $W(\overline{q})$ is a complex $32 \times 32$ -matrix in this case. For a demonstration we display graphically the elements of the $16 \times 16$ -submatrix that corresponds to right movers in fig. \labelcref{fig:U_p_reduced}. The energy eigenstate whose evolution is shown in fig. \labelcref{fig:ism_qm_red_occupationNum_timeEvolution} is one of the energy eigenstates found by this diagonalization. It is a superposition of two different single-orbit states for the orbits which are shown in fig. \labelcref{fig:ism_orbit_structure}.

\section{Density matrix and coarse graining} \label{sect:densityMatrixAndCoarseGraining}

Our setting for the periodic random probabilistic cellular automaton combines randomness and irregularity on short distance scales with regularity on larger distance scales. For the periodic RPCAs the evolution operator is invariant under space-translations by $\Delta x$. It seems therefore natural to \textquote{average out} the short distance irregularity in order to achieve a more regular behavior on some coarse grained level. The quantum formalism offers the appropriate tools for this coarse graining in form of the density matrix. Coarse grained subsystems can be defined by suitable subtraces of the density matrix.

\subsection{Density matrix}

From the complex wave function $\psi_\alpha (x)$ or $\psi_\alpha (q)$ one can form a pure-state density matrix $\rho$ in the standard way, $\alpha, \beta \in \{1, 2\} = \{R, L\}$,
\begin{eq}
    \rho_{\alpha \beta} (x, x') &= \psi_\alpha (x) \psi^*_\beta (x'),\\
    \rho_{\alpha \beta} (q, q') &= \psi_\alpha (q) \psi^*_\beta (q').
\end{eq}
The density matrix in momentum space and position space are related by a discrete Fourier transform
\begin{eq} \label{eq:densityMatrixFourierTransform}
    \rho_{\alpha \beta} (q, q') = N_x^{-1} \sum_{x, x'} \exp\big\{-i(q x - q' x') \rho_{\alpha \beta} (x, x')\big\}.
\end{eq}

The diagonal elements of $\rho(q, q')$ correspond to the probabilities $w(q)$ to find the momentum $q$ (momentum distribution), where $\mathrm{Tr}$ denotes the trace in internal space,
\begin{eq}
    w(q) = \sum_\alpha \rho_{\alpha \alpha} (q, q) = \mathrm{Tr} \rho(q, q).
\end{eq}
As a consequence, the expectation values of functions of momentum obey the quantum rule
\begin{eq}
    \left<f(P)\right> &= \sum_q f(q) w(q) = \sum_q \mathrm{Tr} \rho(q, q) f(q) \\
            &= \tr\{\rho f(P)\}.
\end{eq}
As usual in quantum mechanics, the overall trace $\tr$ can be evaluated in an arbitrary basis.
It is straightforward to derive for classical statistical systems the quantum rule for expectation values
\begin{eq} \label{eq:quantumRuleForExpValues}
    \langle A \rangle (t) = \tr \{\rho(t) \hat{A}\}.
\end{eq}
This holds for all time-local observables that are represented by an operator $\hat{A}$ \cite{CWQMCS, CWEM, CWPT, CWQPCS, CWPW2020}.
The relation \labelcref{eq:quantumRuleForExpValues} demonstrates in a simple way how the quantum rules follow from classical statistics without any additional axioms.

\subsection{Coarse graining in position space}

We label the positions $x$ by a double index $x = (\overline{x}, \xi)$. Here, $\overline{x} = \overline{j} \Delta x$, $\overline{j}$ integer, are the points of the coarse grained lattice and $\xi = h \epsilon, h=0,1,...,(\Delta x/\epsilon) - 1$, labels the positions within an $\Delta x$ interval. Correspondingly, the density matrix takes the form
\begin{eq}
    \rho(x, x') = \rho(\overline{x}, \xi; \overline{x}', \xi'),
    \quad x = \overline{x} + \xi, x' = \overline{x}' + \xi'.
\end{eq}
Coarse graining proceeds by taking the subtrace over the $\xi$-index,
\begin{eq}
    \overline{\rho}(\overline{x}, \overline{x}') = \sum_\xi \rho(\overline{x}+\xi, \overline{x}' + \xi).
\end{eq}
Typically, $\overline{\rho}$ is no longer a pure-state density matrix, i.e. $\overline{\rho}^2 \neq \overline{\rho}$. It remains normalized, however
\begin{eq}
    \tr \overline{\rho} = \sum_{\overline{x}, \alpha} \rho_{\alpha \alpha}(\overline{x}, \overline{x}) = 1.
\end{eq}
We can identify $\rho_{\alpha \alpha} (\overline{x}, \overline{x})$ with the mean occupation number of right-/left movers in the interval $\Delta x$,
\begin{eq}
    \overline{w}_\alpha (\overline{x}) &= \overline{\rho}_{\alpha \alpha} (\overline{x}, \overline{x}),\\
    \left< n_{R/L} (\overline{x}) \right> &= \overline{w}_{R/L} (\overline{x}).
\end{eq}
The occupation number operator $\hat{n}_\gamma (\overline{y})$ for a right-/left moving particle present in the interval $\overline{y} = \overline{j}_y \Delta x$ can take the values one or zero. It is expressed on the coarse grained level by a diagonal operator
\begin{eq}
    (\hat{n}_\gamma (\overline{y}))_{\alpha \beta} (\overline{x}, \overline{x}') = \delta_{\alpha \gamma} \delta_{\alpha \beta} \delta(\overline{y}, \overline{x}) \delta(\overline{x}, \overline{x}').
\end{eq}
Expectation values of observables that are functions of these occupation numbers can be evaluated from the coarse grained density matrix
\begin{eq}
    \left< f(n_\gamma (\overline{y})) \right> = \tr\{\overline{\rho} f(\hat{n}_\gamma (\overline{y}))\},
\end{eq}
where the trace sums over positions of the coarse grained lattice and internal indices.


The notion of a coarse grained wave function is less obvious. First, the coarse grained density matrix does, in general, not correspond to a pure state density matrix, i.e. $\overline{\rho}^2 \neq \overline{\rho}$. If $\overline{\rho}$ is a pure state density matrix one can construct a pure state wave function by the usual procedure of quantum mechanics. Second, the phase information in the wave function may get partially lost in the course of the coarse graining. One may partially overcome these issues by using the density matrix in the real formulation of quantum mechanics. The density matrix becomes then a real symmetric matrix, which contains additional information as compared to the complex density matrix \cite{CWQF, CWPW2020}. The coarse graining can be performed in this real formulation.

\subsection{Coarse grained Fourier transform and momentum}

The coarse grained density matrix can be transformed to Fourier space by a formula similar to eq. \labelcref{eq:densityMatrixFourierTransform}
\begin{eq} \label{eq:coarseGrainedDensityMatrixFourierTransform}
    \overline{\rho}_{\alpha \beta}(\overline{q}, \overline{q}') = \overline{N}_x^{-1} \sum_{\overline{x}, \overline{x}'} \exp[-i (\overline{q}\overline{x} - \overline{q}'\overline{x}')] \overline{\rho}_{\alpha \beta} (\overline{x}, \overline{x}'),
\end{eq}
where $\overline{N}_x = N_x \epsilon / \Delta x$ is the number of $\Delta x$ intervals, e.g. $\overline{j} \in [-\overline{N}_x / 2, \overline{N}_x / 2]$. Correspondingly, $\overline{q}$ or $\overline{q}'$ take $\overline{N}_x$ values,
\begin{eq}
    \overline{q} = \frac{2\pi \overline{k}}{\epsilon N_x}, \quad
    \overline{k} \in \left[-\frac{\overline{N}_x}{2}, \frac{\overline{N}_x}{2}\right],
    \quad \overline{q} \in \left[-\frac{\pi}{\Delta x}, \frac{\pi}{\Delta x}\right],
\end{eq}
with $\overline{k} + \overline{N}_x$ and $\overline{k}$ identified. For large $\Delta x / \epsilon$ the range of $\overline{q}$ is much smaller than the range of $q$. The distance between two neighboring momenta remains the same.

An arbitrary momentum $q$ can be related to the coarse grained momentum $\overline{q}$ by
\begin{eq}
    q = \overline{q} + Q, \quad Q = \frac{2\pi l}{\Delta x},
\end{eq}
where the integer $l$ is in the interval $[-\Delta x / 2\epsilon, \Delta x / 2\epsilon]$. This identifies $\overline{q}$ with $q \mod 2\pi / \Delta x$. As we have seen before, the coarse grained momentum $\overline{q}$ is a conserved quantity.
The associated operator reads in momentum space
\begin{eq}
    \overline{P}(q, q') = \overline{q}(q) \delta_{q, q'}, \quad \overline{q}(q) = q - \frac{2\pi l}{\Delta x}.
\end{eq}
It commutes with the Hamiltonian.

\subsection{Coarse graining in momentum space}

We can define a coarse graining in momentum space by taking for $\rho(q, q')$ a subtrace over Q,
\begin{eq}
    \hat{\rho} (\overline{q}, \overline{q}') = \sum_Q \rho(\overline{q} + Q, \overline{q}' + Q).
\end{eq}
One may be interested how this quantity is related to $\overline{\rho}(\overline{q}, \overline{q}')$ as obtained by a Fourier transform of the coarse grained density matrix in position space. For this purpose we express $\rho(\overline{q} + Q, \overline{q}' + Q)$ in terms of $\rho(x, x')$

\begin{eq} \label{eq:fourierTransformOfQHat}
    \hat{q} (\overline{q}, \overline{q}') = \frac{1}{N_x} \sum_{x, x'} \sum_{Q} &\exp\{-i Q(x - x')\}\\
    &\times \exp\{-i(\overline{q}x - \overline{q}' x')\} \rho(x, x').
\end{eq}
We next employ the relation \labelcref{eq:fourierOrthogonality}
\begin{eq} \label{eq:sumOverExpIsDelta}
    \sum_Q \exp\{-i Q(x - x')\} &= \sum_{l} \exp\{-\frac{2\pi i \epsilon}{\Delta x} l (j - j')\} \\
        &= \frac{\Delta x}{\epsilon} \hat{\delta}(j, j'),
\end{eq}
where $\hat{\delta}(j, j')$ is the delta-function modulo $M_x=\Delta x / \epsilon$. With $\Delta x / (\epsilon N_x) = 1 / \overline{N}_x$ the insertion of eq. \labelcref{eq:sumOverExpIsDelta} into the double sum over $j$ and $j'$ in eq. \labelcref{eq:fourierTransformOfQHat} results in 
\begin{eq}
    &\hat{\rho}(\overline{q}, \overline{q}') = \frac{1}{\overline{N}_x} \sum_{\overline{j}, \overline{j}'} \exp\{-\frac{2\pi i}{\overline{N}_x} (\overline{k} \overline{j} - \overline{k}' \overline{j}')\}\\
        &\times \sum_h \exp\{-\frac{2\pi i}{N_x} (\overline{k} - \overline{k}') h \} \rho(\overline{j} \Delta x + h \epsilon, \overline{j}' \Delta x + h \epsilon),
\end{eq}
where $\overline{x} = \overline{j} \Delta x, \quad x = \overline{j} \Delta x + h \epsilon, \quad \hat{\delta}(j, j') = \delta(h, h')$.
With $\xi = h \epsilon$, we infer
\begin{eq} \label{eq:rhoHatExpressedITORho}
    \hat{\rho}(\overline{q}, \overline{q}') = &\frac{1}{\overline{N}_x} \sum_{\overline{x}, \overline{x}'} \exp\{-i(\overline{q}\overline{x} - \overline{q}' \overline{x}')\}\\
        &\times \sum_\xi \exp\{-i(\overline{q} - \overline{q}')\xi\} \rho(\overline{x} + \xi, \overline{x}' + \xi).
\end{eq}

Comparison with eq. \labelcref{eq:coarseGrainedDensityMatrixFourierTransform} reveals that the diagonal elements of the coarse grained density matrices in momentum spae agree
\begin{eq}
    \overline{\rho}(\overline{q}, \overline{q}') = \hat{\rho}(\overline{q}, \overline{q}).
\end{eq}
Both versions of coarse graining yield the same distribution $\overline{w}(\overline{q}) = \mathrm{Tr} \overline{\rho}(\overline{q}, \overline{q})$ of coarse grained momenta. The off-diagonal elements of $\overline{\rho}$ and $\hat{\rho}$ differ, however, due to the factor $\exp(-i(\overline{q} - \overline{q}')\xi)$ in eq. \labelcref{eq:rhoHatExpressedITORho}. Both ways of coarse graining provide for an easy access to the probabilities for the conserved coarse grained momentum.

\subsection{Coarse grained evolution} \label{sect:coarseGrainedEvolution}

One would like to have an evolution law which formulates quantum mechanics on a coarse grained level. This would combine an averaged evolution in time with some type of averaging in space.
The coarse grained density matrix $\overline{\rho}$ is, however, problematic for a description of an unitary evolution of effective pure states in quantum mechanics. First, $\overline{\rho}$ is, in general, no longer a pure state density matrix -- the relation $\overline{\rho}^2 = \overline{\rho}$ is often not realized even if one starts with a pure state microscopic density matrix $\rho$. Second, the evolution of $\overline{\rho}$ in time is not necessarily unitary.
Information may be exchanged between the subsystem described by $\overline{\rho}$ and its environment. Finally, one may not have direct access to the evolution of $\overline{\rho}$.
An evolution operator $U$ is compatible with the coarse graining if it takes a direct product form, $U = \overline{U} \otimes U_e$, where $\overline{U}$ acts on the coarse grained subsystem and $U_e$ influences only the environment. In this case the time evolution of $\overline{\rho}$ is described by a unitary evolution encoded in $\overline{U}$. For the coarse graining in position or momentum space described above this direct product property is not realized.
In this case the coarse grained density matrix is mainly an analysis tool, rather than being used for a coarse gained evolution law. 

The main idea of some type of coarse grained evolution is to get rid of fast oscillations in time. If a measurement device involves a typical time scale $\Delta \tau$ it cannot resolve the oscillations of the wave function or density matrix on time scales much smaller than $\Delta \tau$. One somehow wants to \textquote{integrate out} or \textquote{remove} the fast oscillations. If the energy spectrum has a clear separation between a sector of \textquote{small energies} and \textquote{high energies}, one may discard the fast oscillations associated to the high energies by restricting the wave function to linear combinations of eigenfunctions for the small energies. In particle physics this corresponds to the concept of an effective low energy theory.
\textquote{Integrating out the heavy particles} in particle physics can be seen as a procedure to make the Hamiltonian block diagonal in the small and large energies.

For the periodic RPCAs one may guess a separation between small and large energies if $N_x$ gets very large and coarse grained momentum scales as $\overline{p} \sim 2\pi / N_x$, with other parameters fixed. Indeed, for the periodic random automaton with $\eta = 1$ one may expect that for small $\overline{q}$ there exists an energy $E(\overline{q})$ which vanishes for $\overline{q} \rightarrow 0$, either $\sim \overline{q}$ or even $\sim\overline{q}^2$. This is motivated by the exact result that an eigenvalue $E=0$ exists for $\overline{q} = 0$.
It corresponds to the static space-independent solution. For a given $\overline{q}$ the other energy eigenvalues may be distributed over the available energy interval $|E| \leq \pi / \Delta t$. This distribution has typical distances between energy levels $\Delta E \sim \pi / (2M_x \Delta t)$.
We could identify these other energy eigenvalues with the large energies. If we want to achieve a clear separation between small and large energies with $\overline{q} \ll \Delta E$ in the range where $\overline{q} \sim 2\pi / (\epsilon N_x)$, we need
\begin{eq}
    N_x \gg 4 M_x M_t = \frac{4 \Delta x \Delta t}{\epsilon^2}.
\end{eq}
For numerical simulations this needs a substantial number of lattice points even for moderate $\Delta x / \epsilon$ and $\Delta t / \epsilon$.

We have found these properties for the periodic RPCAs for which we have diagonalized the Hamiltonian for given $\overline{q}$. These systems admit a family of small energy eigenvalues $E(\overline{q})$ which vanish for $\overline{q} \rightarrow 0$. Since these systems have rather moderate $\Delta x / \epsilon$ and $\Delta t / \epsilon$ and rare scattering points, the eigenstates of the Hamiltonian are found to be single-orbit states or combinations of single-orbit states for orbits with the same length. Correspondingly, we have found a linear dispersion relation \labelcref{eq:linearDispersionRelation}\labelcref{eq:linearDispersion_pc_E}.
As we have argued above, the linearity of the dispersion relation is not expected to be maintained in the continuum limit.

\section{Continuum limit} \label{sect:continuumLimit}

A standard continuum limit may be realized by the limit $\epsilon \rightarrow 0$, or equivalently $N_x \rightarrow \infty$, for fixed $L$.
If we also hold \textquote{physical time intervals} fixed, the continuum limit likewise extrapolates the number of time steps to infinity. Increasing, on the other hand, the number of lattice sites $N_x \rightarrow \infty$ while keeping $\epsilon$ fixed corresponds to the \textquote{infinite volume limit} $L \rightarrow \infty$.
A smooth continuum limit requires the initial wave function to be sufficiently smooth, and that this property is maintained during the evolution in time.
An approach to the infinite volume limit allows us to explore smaller and smaller (coarse grained) momenta $p$.
Plane waves with small nonzero $|p|$ are close to the static homogeneous solution for $p=0$ which exists for $\eta=1$. For these states with $\phi_R(x) \approx \phi_L(x)$ an individual scattering induces only a small change in the wave function. Also the transport by the free part of the step evolution operator results only in a small change of the wave function. These are the ingredients for the possible realization of a naive continuum limit. For the random automaton, one may speculate that this smoothness extends to the energy eigenstates with small energies.

If one multiplies $N_x$ and $L$ by an integer factor $l$, keeping $\Delta x$ fixed, the length of full orbits changes from $N_m$ to $l N_m$ for all orbits with non-zero winding number. From the point of view of the extended system the previous orbits with length $N_m$ can be interpreted as reduced orbits. For a single-orbit state the energy levels \labelcref{eq:singleOrbitEnergies} get denser, replacing $N_m$ by $l N_m$.
For two orbits with different length $N_1 \neq N_2$ the single-orbit states have common energy eigenvalues if there exist integers $k_1$ and $k_2$ such that
\begin{eq}
    \frac{k_1}{N_1 l} = \frac{k_2}{N_2 l}, \quad k_2 = \frac{N_2}{N_1}k_1.
\end{eq}
Since the allowed range of $k_1$ and $k_2$ increases $\sim l$, more and more energy levels are shared by the two single-orbit states. This underlines our general discussion above, that in the continuum limit the mixing of single-orbit states becomes generic for simultaneous eigenstates of energy and momentum.

Even for large $N_x$ (or $l N_x$) the existence of smooth eigenfunctions for the small energies is not guaranteed, however. Smoothness of the eigenfunctions may require, in addition, large $\Delta t / \epsilon$, or it may not be realized at all for the RPCAs investigated in this note.

\subsection{Evolution of plane waves}

As a first step towards a possible continuum limit we start with an initial plane wave. Since this is not an energy eigenstate of the RPCA the subsequent evolution may drive the wave function away from an approximate plane wave form. Still, for a certain time $t_p$ the plane wave may remain a good approximation of the wave function, and one may investigate a possible continuum limit which is valid for $t < t_p$.

Consider first the action of $U(0) = U_s(0) U_f$ on the initial plane wave \labelcref{eq:plainWave}
\begin{eq}
    \psi_p (0, q) = \begin{pmatrix}f(q)\\i f(-q)\end{pmatrix} \delta_{p, q}.
\end{eq}
The value of $m$ in $f(q)$, eq. \labelcref{eq:plainWaveParameter}, is arbitrary at this stage and may be chosen self-consistently later. With
\begin{eq}
    U_f = \exp\{-i \epsilon P \tau_3 \}
\end{eq}
the first factor yields
\begin{eq}
    U_f \psi_p (0, q) = \begin{pmatrix}e^{-i \epsilon q} f(q)\\ i e^{i \epsilon q} f(-q)\end{pmatrix} \delta_{p, q} = \psi_p (0, q) + \tilde{\delta} \psi_p (q).
\end{eq}
In lowest order of an expansion in small $\epsilon p$ one finds for the small change of the wave function
\begin{eq}
    \tilde{\delta} \psi_p (q) = -i \epsilon p \begin{pmatrix}f(p)\\-i f(-p)\end{pmatrix} \delta_{p, q}.
\end{eq}

We next turn to the action of the second factor $U_s$ which is best studied in position space.
In position space the initial plane wave reads
\begin{eq}
    \psi_p (0, x) = N_x ^{-1/2} \exp(i p x) \begin{pmatrix}f(p)\\i f(-p)\end{pmatrix},
\end{eq}
and the action of $U_s(0)$ on this plane wave yields
\begin{eq}
    U_s(0) \psi_p (0,x) = &\psi_p (0, x) + \hat{\delta} \psi_p (x),
\end{eq}
with
\begin{eq}
    \hat{\delta} \psi_p (x) = &N_x^{-1/2} \sum_j \exp\{i p \overline{x}_j (0)\} \\
    &\times (\eta \tau_2 - 1)\begin{pmatrix}f(p)\\i f(-p)\end{pmatrix} \delta_{x, \overline{x}_{j}(0)}.
\end{eq}
For $\eta=1$ one finds
\begin{eq}
    (\tau_2 - 1)\begin{pmatrix}f(p)\\i f(-p)\end{pmatrix} = -(f(p) - f(-p)) \begin{pmatrix}1\\-i\end{pmatrix},
\end{eq}
which results for $p^2 \ll m^2$ in
\begin{eq}
    \hat{\delta} \psi_p (x) = -(2N_x)^{-1/2} \frac{p}{m} \sum_j \exp\{i p \overline{x}_j (0)\} \begin{pmatrix}1\\-i\end{pmatrix} \delta_{x, \overline{x}_j (0)}.
\end{eq}
One concludes that for small $|p/m|$ and $|\epsilon p|$ both $U_f$ and $U_s(0)$ only induce a small change of the plane wave. This does not hold for $\eta = -i$.

The action of $U_s(0)$ on $\tilde{\delta} \psi_p$ results in
\begin{eq}
    U_s(0) \tilde{\delta}\psi_p (x) = &-i \epsilon p N_x^{-1/2} \biggl[ e^{i p x} \begin{pmatrix}f(p)\\-i f(-p)\end{pmatrix} \\
    &- \sqrt{2} \sum_j e^{i p \overline{x}_j(0)} \begin{pmatrix}1\\-i\end{pmatrix} \delta_{x, \overline{x}_j (0)}\biggr].
\end{eq}
For the product $U_s(0) U_f$ we therefore obtain the leading contribution for small $|p|/m$ and $\epsilon |p|$
\begin{eq}
    U(0) \psi_p (x) = \psi_p (x) + \delta \psi_p (x),
\end{eq}
with
\begin{eq} \label{eq:momentumExpansion}
    \delta \psi_p (x) = &\hat{\delta} \psi_p (x) + U_s (0) \tilde{\delta} \psi_p (x)\\
    = &- \frac{i \epsilon p^2}{2m} \psi_p (0, x) -\frac{1}{\sqrt{2N_x}} \bigg\{ i \epsilon p e^{i p x} \\
      &+ \frac{p}{m} (1 - 2 i \epsilon m) \sum_j e^{i p \overline{x}_j (0)} \delta_{x, \overline{x}_j (0)}\bigg\} \begin{pmatrix}1\\-i\end{pmatrix}. \\
\end{eq}
In the limit where the curly bracket can be neglected this amounts to the evolution for a Dirac fermion with mass $m$ in the non-relativistic limit.

\subsection{Generalized Potential}

The scattering part $U_s(t)$ can be represented in terms of a matrix valued generalized potential $\overline{V}(t, x)$
\begin{eq}
    U_s(t; x, x') &= \exp\{-i \epsilon \overline{V}(t,x)\} \delta_{x, x'},\\
    \overline{V}(t,x) &= \frac{\pi}{2\epsilon} (\tau_2 - c) \sum_j \delta_{x, \overline{x}_j (t)}.
\end{eq}
Here $\overline{x}_j(t)$ denote the positions of the scattering points at a given $t$, and $c=0$ for $\eta=-i$ and $c=1$ for $\eta=1$. In terms of this potential we write in position space
\begin{eq}
    U(0) = \exp \{ -i \epsilon \overline{V}(0,x) \} \exp \{ -i \epsilon P \tau_3 \} = \exp \{ -i \epsilon \overline{H}(0) \},
\end{eq}
with
\begin{eq}
    \overline{H}(0) = P \tau_3 + \overline{V}(0, x) + O(\epsilon[P, \overline{V}]).
\end{eq}

Since $\overline{V}$ contains a factor $1/\epsilon$ it is a priori not clear if the commutator correction vanishes for $\epsilon \rightarrow 0$. This is an important difference between the random automaton and the Dirac particle. For the latter one has $\overline{V} = m (\tau_2 - 1)$, such that the commutator term indeed vanishes for $\epsilon \rightarrow 0$.
In order to estimate the role of the commutator term for the random automaton, we compute
\begin{eq}
    \exp \{ -i \epsilon (P \tau_3 + \overline{V}(0, x)) \} \psi_p (0, x) \\
    = \psi_p (0, x) + \delta_{0} \psi_p (x).
\end{eq}
For $\eta = 1$ one finds for $\delta_0 \psi_p (x)$ the expression \labelcref{eq:momentumExpansion} for $\delta \psi_p (x)$ with the modification that one omits in the term $\sim \delta_{x, \overline{x}_j (0)}$ the factor $(1 - 2i \epsilon m)$.
We conclude that for small $\epsilon m$ the correction from the commutator term is proportional to $\epsilon m$ and therefore vanishes for $\epsilon m \rightarrow 0$.

The time $t=0$ is not particularly singled out and our discussion applies to arbitrary $t$ as long as the wave function remains close to a plane wave with $\epsilon m \ll 1, |\epsilon p | \ll 1$. In this case the step evolution operator is approximated by a microscopic Hamiltonian for a quantum particle with a particular time- and space- dependent matrix-potential.

\subsection{Mesoscopic evolution and naive continuum limit}

The mesoscopic evolution operator \labelcref{eq:mesoscopicEvolutionOperator} involves an ordered product
\begin{eq}
    \exp(-i \Delta t H) = e^{-i \epsilon \overline{H}(t + \Delta t - \epsilon)} ... e^{-i \epsilon \overline{H}(t + \epsilon)} e^{-i \epsilon \overline{H}(t)}.
\end{eq}
The naive continuum limit neglects the non-vanishing commutators $[\overline{H}(t_1), \overline{H}(t_2)]$ and approximates $\overline{H}(t) = P \tau_3 + \overline{V}(t, x)$.
In this approximation one obtains ($\eta = 1$)
\begin{eq}
    H &= P \tau_3 + \frac{\epsilon}{\Delta t} \sum_t \overline{V}(t, x) \\
      &= P \tau_3 + \frac{\pi}{2 \Delta t} (\tau_2 - 1) \sum_t \sum_j \delta_{x, \overline{x}_j (t)} \\
      &= P \tau_3 + M(x) (\tau_2 - 1).
\end{eq}
At any given position $x$ the quantity $M(x)$ is proportional to the number of scattering points at $x$ within the interval $\Delta t$.

Neglecting the fluctuations around the mean value of scattering points per site we may approximate
\begin{eq}
    \sum_t \sum_j \delta_{x, \overline{x}_j (t)} \approx \overline{n} \Delta t / \epsilon.
\end{eq}
Here we recall that $\overline{n} = n_{tot} \frac{\epsilon^2}{\Delta x \Delta t}$ is the mean number of scattering points at a given $x$ for a given $t$. In this approximation one finds
\begin{eq}
    M(x) = \overline{M} = \frac{\pi \overline{n}}{2\epsilon}.
\end{eq}
We can now choose self-consistently $m = \overline{M}$. For this approximation the naive continuum limit is identical to the discrete quantum mechanics for the Dirac particle, with
\begin{eq} \label{eq:massScatteringDensity}
    m \epsilon = \frac{\pi \overline{n}}{2} = \frac{\pi n_{tot} \epsilon^2}{2 \Delta x \Delta t}.
\end{eq}
The inhomogeneity in $M(x)$ can be avoided if we restrict the distribution of scattering points in the interval $\Delta t \Delta x$ such that every position $x$ receives an equal number of points. More generally, an inhomogeneous distribution can realize a potential $V(x)$ for the quantum particle \cite{CWPCAQP}.

The validity of the naive continuum limit depends critically on the neglection of commutator terms $[\overline{H}(t_1), \overline{H}(t_2)]$ for $t_1 \neq t_2$. It may hold approximately as long as the wave function remains sufficiently smooth.
As long as the wave function can be approximated by a plane wave we may employ ($\eta = 1$)

\begin{eq}
    \overline{H}(t) &= P \tau_3 + \overline{V}(t, x) = P \tau_3 + \tilde{M}(t, x) (\tau_2 - 1),\\
    \tilde{M}(t, x) &= \frac{\pi}{2\epsilon} \sum_j \delta_{x, \overline{x}_j} (t),
\end{eq}
with the commutator
\begin{eq}
    \Delta(t_1, t_2) &= \epsilon [\overline{H}(t_1), \overline{H}(t_2)]\\ 
    &= [P \tau_3, \epsilon (\tilde{M}(t_2, x) - \tilde{M}(t_1, x))(\tau_2 - 1)]\\
    &= [P \tau_3, D(x)(\tau_2 - 1)]\\
    &= -[P, D]\tau_3 - i \{P, D\}\tau_1,
\end{eq}
where
\begin{eq}
    D(x) = \frac{\pi}{2} \sum_j (\delta_{x, \overline{x}_j(t_2)} - \delta_{x, \overline{x}_j (t_1)}).
\end{eq}
For the validity of the naive continuum limit the action of an appropriate sum of terms $\Delta (t_a, t_b)$ on the wave function has to be small as compared to the one of $H = P \tau_3 + M(\tau_2 - 1)$\cite{CWFPPCA, CWPCAQP}.

\subsection{Dispersion relation}

For the Dirac particle with $m \epsilon \ll 1, p \epsilon \ll 1$ the plane wave solutions are eigenstates of both momentum and energy. The dispersion relation between energy and momentum reads
\begin{eq} \label{eq:dispersionRelation}
    E(p) = \sqrt{p^2 + m^2} - m.
\end{eq}
One could check the possible validity of the naive continuum limit for the random automaton by establishing the dispersion relation $E(p) = E(\overline{q}=p)$ for the energy eigenstates with small $|E|$.

\begin{figure}[t]\centering
    \includegraphics[width=8.5cm]{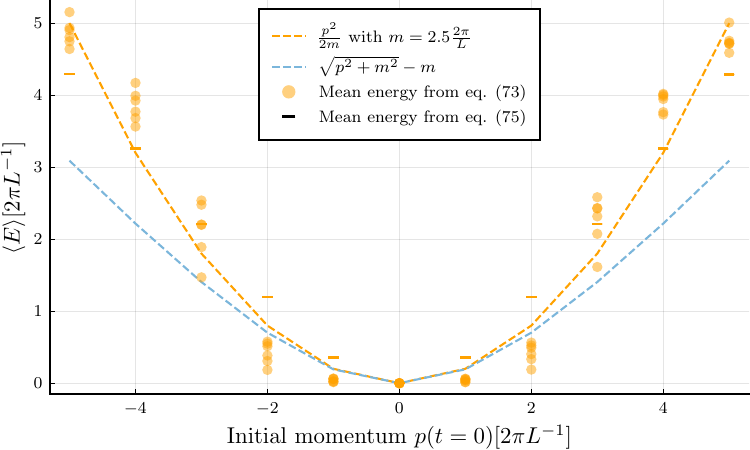}
    \caption{Mean energy for plane wave initial states with different momenta for the Brownian automaton \labelcref{eq:brownianModelParams}, however with $N_x=4096, m = 20 \cdot 2\pi / L$, which corresponds to an approaching of the large volume limit. The different colors show different automata, corresponding to different distributions of the randomly chosen scattering points. The continuous curve indicates the dispersion relation for the Dirac particle with mass $m=2.5 (2\pi / L)$, corresponding to the naive continuum limit for the Brownian automaton. For the initial stages of the evolution for $\overline{N}_t = 64 \Delta t$-intervals the naive continuum limit appears as a rough approximation.}
    \label{fig:periodicBrownianModel_dispersion}
\end{figure}

If the energy eigenstates of the automaton are not known explicitly, as for the Brownian automaton, one may still get a rough impression by plotting the mean energy $\langle E \rangle$ as a function of the momentum of the initial plane wave. This is done in fig. \labelcref{fig:periodicBrownianModel_dispersion}. For small $p$ the dispersion relation for the Dirac fermion becomes quadratic, $E(p) \approx p^2 / (2m)$. Comparing $\langle E \rangle (p)$ for the Brownian automaton with the dispersion relation \labelcref{eq:dispersionRelation} for the Dirac fermion one finds rough agreement.

The inhomogeneity of the scattering points may be incorporated by a mean number of scattering points $\overline{n}$ which corresponds to the mass of the continuum limit by eq. \labelcref{eq:massScatteringDensity}, and a deviation of the actual distribution from this mean number. Qualitatively, this may correspond to a massive Dirac fermion in a random potential. The quantum mechanics of a randomly scattered Dirac fermion may be closer to the Brownian automaton. It would be interesting to see if and how a continuum limit for the randomly scattered Dirac fermion is reached.

\section{Discussion and conclusions}\label{sect:discussion}

In this paper we have demonstrated how the methods of quantum mechanics provide insight into the evolution of probabilistic cellular automata with deterministic updating. On the one side these probabilistic automata, as characterized by a probability distribution for the initial configurations, are classical statistical systems. On the other side, casting the probabilistic information into the form of (\textquote{classical} or \textquote{real}) wave functions reveals that probabilistic automata are quantum systems. This equivalence demonstrates how quantum systems emerge as special cases of classical statistical systems \cite{CWQMCS, CWEM, CWPT, CWQPCS, wetterich_classical_2011}.

The description in terms of wave functions exhibits some redundancy in form of a local discrete gauge symmetry associated to signs of the real wave function. For a fixed gauge, as used here, the sign convention is fixed and plays no further role. For all observables that take definite values for given bit-configuration of the automaton the expectation values can be found by an updating of the probability distribution, without the use of wave functions.
The formulation in terms of wave functions provides, however, powerful tools for the understanding of the evolution of the probabilistic information.
One can rely on the whole apparatus of quantum mechanics.

A complex structure can be implemented for many probabilistic automata. It only requires the presence of two discrete symmetries, corresponding to complex conjugation and multiplication by $i$, which have to be compatible with the updating law of the automaton\cite{CWQMCS}. In the present paper we realize a rather trivial complex structure associated to the two colors of the bits.
In the presence of a complex structure the probabilistic automaton obeys all usual rules of quantum mechanics, with complex wave functions spanning a complex Hilbert space, Hermitian operators for observables and a complex discrete Schrödinger equation.

Central quantities for an understanding of the time evolution of the probabilistic information are energy and momentum. Energy eigenstates correspond to probability distributions which are periodic in time.
The generic existence of a very large number of periodic distributions for random probabilistic automata would be rather difficult to infer from a description based solely on the probability distribution. (The particular special cases of single-orbit states are an exception.)
Momentum eigenstates show periodicity in space. As familiar from quantum mechanics the conservation of energy or momentum corresponds to translation invariance in time or space, expressed by the vanishing commutator of the associated operators with the Hamiltonian.

For the very simple automata described in this paper the complex wave function corresponds to a single two-component quantum particle in one space- and one time-dimension.
Together with the momentum operator taking the same form as for discrete one-particle quantum mechanics we observe a strong formal analogy to one-particle quantum mechanics.
Only the Hermitian Hamiltonian does not take the standard form for a particle propagating in a potential. One of our aims is the characterization of the properties of the Hamiltonian for the investigated random probabilistic automaton. We have developed analysis tools based on the discrete Fourier transform to frequency space for the transition element, determination of expectation value and variance of the energy observable, or diagonalization of the evolution operator in momentum space. These methods rely crucially on the quantum mechanical description, in particular the possibility to perform basis transformations for the wave function.
For very simple single-orbit states we have constructed the Hamilton operator and its eigenstates explicitly.

As steps towards a possible continuum limit towards an infinite number of cells we discuss coarse graining in position and momentum space, as well as the notion of a \textquote{low energy effective theory} which focuses on linear superpositions of eigenstates to \textquote{low energy eigenvalues}. Applied on smooth initial plane waves the Hamiltonian of the random probabilistic automaton is found to be close to the Hamiltonian for a free massive non-relativistic particle. Correspondingly, the early evolution of the wave function is close to the one for a free massive particle in quantum mechanics.
For the limited number of cells and scattering points considered in the present paper the wave function for an initial plane wave starts to become rough after a certain time. A full continuum limit has not been found in this case.
As momenta are lowered in the infinite volume limit the differences between different realizations of the Brownian automaton become smaller.
As a future way towards a smooth continuum limit one should look for initial states that are closer to eigenstates of the Hamilton operator.
Without approaching the energy eigenstates, the number of visible periods in the evolution does not increase, such that the dispersion apparent in fig. \labelcref{fig:periodicBrownianModel_dispersion} remains a short term phenomenon.
At present, it is not known if the particular stochastic probabilistic automata considered in the present paper admit a smooth continuum limit, or if some roughness of the energy eigenfunctions remains even in the continuum limit.

For the description of a quantum particle in a potential by a probabilistic automaton \cite{CWPW2020, CWFPPCA, CWPCAQP} it may be necessary to explore more complex automata beyond our setting with a single occupied bit.
One may have to consider an arbitrary number of occupied bits, and start with half-filled vacua with all negative-energy-states filled \cite{CWPW2020}.
As familiar for quantum field theories of fermions the one-particle states can then be defined as excitations of such a vacuum.
Needless to say that a numerical investigation of this setting would require substantially larger resources, even for a rather moderate number of cells.
As a possible intermediate step one may extend the one-bit automaton by introducing the evolution of additional bits which could mimic the presence of a non-trivial vacuum. This could help to smoothen energy eigenfunctions and to allow for an easier access to a continuum limit.

In any case, the probabilistic automata with deterministic updating are interesting systems in their own right. The availability of the full formalism of quantum mechanics opens many new avenues for the understanding of their evolution.
On the conceptual level we have demonstrated both that quantum mechanics constitutes a particular case of classical statistical evolution, and that the quantum formalism with wave functions and operators is a useful tool for the understanding of the evolution of classical statistical systems.

\section*{Acknowledgement}

This work has been supported by the DFG collaborative research center SFB 1225 ISOQUANT and by the DFG excellence cluster \textquote{STRUCTURES}.


\nocite{*} 
\bibliography{refs}

\begin{thebibliography}{53}%
\makeatletter
\providecommand \@ifxundefined [1]{%
 \@ifx{#1\undefined}
}%
\providecommand \@ifnum [1]{%
 \ifnum #1\expandafter \@firstoftwo
 \else \expandafter \@secondoftwo
 \fi
}%
\providecommand \@ifx [1]{%
 \ifx #1\expandafter \@firstoftwo
 \else \expandafter \@secondoftwo
 \fi
}%
\providecommand \natexlab [1]{#1}%
\providecommand \enquote  [1]{``#1''}%
\providecommand \bibnamefont  [1]{#1}%
\providecommand \bibfnamefont [1]{#1}%
\providecommand \citenamefont [1]{#1}%
\providecommand \href@noop [0]{\@secondoftwo}%
\providecommand \href [0]{\begingroup \@sanitize@url \@href}%
\providecommand \@href[1]{\@@startlink{#1}\@@href}%
\providecommand \@@href[1]{\endgroup#1\@@endlink}%
\providecommand \@sanitize@url [0]{\catcode `\\12\catcode `\$12\catcode `\&12\catcode `\#12\catcode `\^12\catcode `\_12\catcode `\%12\relax}%
\providecommand \@@startlink[1]{}%
\providecommand \@@endlink[0]{}%
\providecommand \url  [0]{\begingroup\@sanitize@url \@url }%
\providecommand \@url [1]{\endgroup\@href {#1}{\urlprefix }}%
\providecommand \urlprefix  [0]{URL }%
\providecommand \Eprint [0]{\href }%
\providecommand \doibase [0]{http://dx.doi.org/}%
\providecommand \selectlanguage [0]{\@gobble}%
\providecommand \bibinfo  [0]{\@secondoftwo}%
\providecommand \bibfield  [0]{\@secondoftwo}%
\providecommand \translation [1]{[#1]}%
\providecommand \BibitemOpen [0]{}%
\providecommand \bibitemStop [0]{}%
\providecommand \bibitemNoStop [0]{.\EOS\space}%
\providecommand \EOS [0]{\spacefactor3000\relax}%
\providecommand \BibitemShut  [1]{\csname bibitem#1\endcsname}%
\let\auto@bib@innerbib\@empty
\bibitem [{\citenamefont {von Neumann}(1951)}]{JVN}%
  \BibitemOpen
  \bibfield  {author} {\bibinfo {author} {\bibfnamefont {John}\ \bibnamefont {von Neumann}},\ }\enquote {\bibinfo {title} {The general and logical theory of automata.}}\ \ (\bibinfo  {publisher} {Wiley},\ \bibinfo {address} {Oxford, England},\ \bibinfo {year} {1951})\ pp.\ \bibinfo {pages} {1--41}\BibitemShut {NoStop}%
\bibitem [{\citenamefont {Ulam}(1950)}]{ULA}%
  \BibitemOpen
  \bibfield  {author} {\bibinfo {author} {\bibfnamefont {S.}~\bibnamefont {Ulam}},\ }\bibfield  {title} {\enquote {\bibinfo {title} {Random processes and transformations},}\ }in\ \href@noop {} {\emph {\bibinfo {booktitle} {Proceedings of the International Congress on Mathematics}}},\ Vol.~\bibinfo {volume} {2}\ (\bibinfo {year} {1950})\ p.\ \bibinfo {pages} {264–275}\BibitemShut {NoStop}%
\bibitem [{\citenamefont {Zuse}(1969)}]{ZUS}%
  \BibitemOpen
  \bibfield  {author} {\bibinfo {author} {\bibfnamefont {Konrad}\ \bibnamefont {Zuse}},\ }\href@noop {} {\emph {\bibinfo {title} {Rechnender Raum}}}\ (\bibinfo  {publisher} {Vieweg, Teubner Verlag},\ \bibinfo {year} {1969})\ p.\ \bibinfo {pages} {1–3}\BibitemShut {NoStop}%
\bibitem [{\citenamefont {Gardner}(1970)}]{GAR}%
  \BibitemOpen
  \bibfield  {author} {\bibinfo {author} {\bibfnamefont {Martin}\ \bibnamefont {Gardner}},\ }\bibfield  {title} {\enquote {\bibinfo {title} {Mathematical games},}\ }\href@noop {} {\bibfield  {journal} {\bibinfo  {journal} {Scientific American}\ }\textbf {\bibinfo {volume} {223}},\ \bibinfo {pages} {120–123} (\bibinfo {year} {1970})}\BibitemShut {NoStop}%
\bibitem [{\citenamefont {Lindenmayer}\ and\ \citenamefont {Rozenberg}(1976)}]{LIRO}%
  \BibitemOpen
  \bibfield  {author} {\bibinfo {author} {\bibfnamefont {Aristid}\ \bibnamefont {Lindenmayer}}\ and\ \bibinfo {author} {\bibfnamefont {Grzegorz}\ \bibnamefont {Rozenberg}},\ }\bibfield  {title} {\enquote {\bibinfo {title} {Automata, languages, development},}\ \ }(\bibinfo  {publisher} {North Holland},\ \bibinfo {year} {1976})\BibitemShut {NoStop}%
\bibitem [{\citenamefont {Toom}(1978)}]{TOOM}%
  \BibitemOpen
  \bibfield  {author} {\bibinfo {author} {\bibfnamefont {A.~L.}\ \bibnamefont {Toom}},\ }\href@noop {} {\emph {\bibinfo {title} {Locally Interacting Systems and their application in Biology}}}\ (\bibinfo  {publisher} {Springer Berlin Heidelberg},\ \bibinfo {year} {1978})\BibitemShut {NoStop}%
\bibitem [{\citenamefont {{R. L. Dobrushin}}(1978)}]{DKT}%
  \BibitemOpen
  \bibfield  {author} {\bibinfo {author} {\bibfnamefont {A.~L.~Toom}\ \bibnamefont {{R. L. Dobrushin}}, \bibfnamefont {V.I.~Kryukov}},\ }\href@noop {} {\emph {\bibinfo {title} {Stochastic cellular systems: Ergodicity, Memory, Morphogenesis}}}\ (\bibinfo  {publisher} {Manchester University Press},\ \bibinfo {year} {1978})\BibitemShut {NoStop}%
\bibitem [{\citenamefont {Wolfram}(1983)}]{WOLF}%
  \BibitemOpen
  \bibfield  {author} {\bibinfo {author} {\bibfnamefont {Stephen}\ \bibnamefont {Wolfram}},\ }\bibfield  {title} {\enquote {\bibinfo {title} {Statistical mechanics of cellular automata},}\ }\href {\doibase 10.1103/RevModPhys.55.601} {\bibfield  {journal} {\bibinfo  {journal} {Rev. Mod. Phys.}\ }\textbf {\bibinfo {volume} {55}},\ \bibinfo {pages} {601–644} (\bibinfo {year} {1983})}\BibitemShut {NoStop}%
\bibitem [{\citenamefont {Vichniac}(1984)}]{VICH}%
  \BibitemOpen
  \bibfield  {author} {\bibinfo {author} {\bibfnamefont {Gérard~Y.}\ \bibnamefont {Vichniac}},\ }\bibfield  {title} {\enquote {\bibinfo {title} {Simulating physics with cellular automata},}\ }\href@noop {} {\bibfield  {journal} {\bibinfo  {journal} {Physica D: Nonlinear Phenomena}\ }\textbf {\bibinfo {volume} {10}},\ \bibinfo {pages} {96–116} (\bibinfo {year} {1984})}\BibitemShut {NoStop}%
\bibitem [{\citenamefont {Preston}\ and\ \citenamefont {Duff}(1984)}]{PREDU}%
  \BibitemOpen
  \bibfield  {author} {\bibinfo {author} {\bibfnamefont {Kendall}\ \bibnamefont {Preston}}\ and\ \bibinfo {author} {\bibfnamefont {Michael J.~B.}\ \bibnamefont {Duff}},\ }\href@noop {} {\emph {\bibinfo {title} {Modern Cellular Automata}}}\ (\bibinfo  {publisher} {Springer {US}},\ \bibinfo {year} {1984})\ p.\ \bibinfo {pages} {1–15}\BibitemShut {NoStop}%
\bibitem [{\citenamefont {Toffoli}\ and\ \citenamefont {Margolus}(1990)}]{TOMA}%
  \BibitemOpen
  \bibfield  {author} {\bibinfo {author} {\bibfnamefont {Tommaso}\ \bibnamefont {Toffoli}}\ and\ \bibinfo {author} {\bibfnamefont {Norman~H.}\ \bibnamefont {Margolus}},\ }\bibfield  {title} {\enquote {\bibinfo {title} {Invertible cellular automata: A review},}\ }\href@noop {} {\bibfield  {journal} {\bibinfo  {journal} {Physica D: Nonlinear Phenomena}\ }\textbf {\bibinfo {volume} {45}},\ \bibinfo {pages} {229–253} (\bibinfo {year} {1990})}\BibitemShut {NoStop}%
\bibitem [{\citenamefont {LOUIS}\ and\ \citenamefont {Nardi}(2018)}]{FLN}%
  \BibitemOpen
  \bibfield  {author} {\bibinfo {author} {\bibfnamefont {Pierre-Yves}\ \bibnamefont {LOUIS}}\ and\ \bibinfo {author} {\bibfnamefont {Francesca~R.}\ \bibnamefont {Nardi}},\ }\href@noop {} {\emph {\bibinfo {title} {{Probabilistic Cellular Automata}}}}\ (\bibinfo  {publisher} {Springer},\ \bibinfo {year} {2018})\BibitemShut {NoStop}%
\bibitem [{\citenamefont {Hedlund}(1969)}]{HED}%
  \BibitemOpen
  \bibfield  {author} {\bibinfo {author} {\bibfnamefont {G.~A.}\ \bibnamefont {Hedlund}},\ }\bibfield  {title} {\enquote {\bibinfo {title} {Endomorphisms and automorphisms of the shift dynamical system},}\ }\href@noop {} {\bibfield  {journal} {\bibinfo  {journal} {Mathematical systems theory}\ }\textbf {\bibinfo {volume} {3}},\ \bibinfo {pages} {320--375} (\bibinfo {year} {1969})}\BibitemShut {NoStop}%
\bibitem [{\citenamefont {Richardson}(1972)}]{RICH}%
  \BibitemOpen
  \bibfield  {author} {\bibinfo {author} {\bibfnamefont {D.}~\bibnamefont {Richardson}},\ }\bibfield  {title} {\enquote {\bibinfo {title} {Tessellations with local transformations},}\ }\href@noop {} {\bibfield  {journal} {\bibinfo  {journal} {Journal of Computer and System Sciences}\ }\textbf {\bibinfo {volume} {6}},\ \bibinfo {pages} {373–388} (\bibinfo {year} {1972})}\BibitemShut {NoStop}%
\bibitem [{\citenamefont {Amoroso}\ and\ \citenamefont {Patt}(1972)}]{AMPA}%
  \BibitemOpen
  \bibfield  {author} {\bibinfo {author} {\bibfnamefont {S.}~\bibnamefont {Amoroso}}\ and\ \bibinfo {author} {\bibfnamefont {Y.N.}\ \bibnamefont {Patt}},\ }\bibfield  {title} {\enquote {\bibinfo {title} {Decision procedures for surjectivity and injectivity of parallel maps for tessellation structures},}\ }\href@noop {} {\bibfield  {journal} {\bibinfo  {journal} {Journal of Computer and System Sciences}\ }\textbf {\bibinfo {volume} {6}},\ \bibinfo {pages} {448–464} (\bibinfo {year} {1972})}\BibitemShut {NoStop}%
\bibitem [{\citenamefont {Hardy}\ \emph {et~al.}(1976)\citenamefont {Hardy}, \citenamefont {de~Pazzis},\ and\ \citenamefont {Pomeau}}]{HPP}%
  \BibitemOpen
  \bibfield  {author} {\bibinfo {author} {\bibfnamefont {J.}~\bibnamefont {Hardy}}, \bibinfo {author} {\bibfnamefont {O.}~\bibnamefont {de~Pazzis}}, \ and\ \bibinfo {author} {\bibfnamefont {Y.}~\bibnamefont {Pomeau}},\ }\bibfield  {title} {\enquote {\bibinfo {title} {Molecular dynamics of a classical lattice gas: Transport properties and time correlation functions},}\ }\href@noop {} {\bibfield  {journal} {\bibinfo  {journal} {Phys. Rev. A}\ }\textbf {\bibinfo {volume} {13}},\ \bibinfo {pages} {1949–1961} (\bibinfo {year} {1976})}\BibitemShut {NoStop}%
\bibitem [{\citenamefont {Creutz}(1986)}]{CREU}%
  \BibitemOpen
  \bibfield  {author} {\bibinfo {author} {\bibfnamefont {Michael}\ \bibnamefont {Creutz}},\ }\bibfield  {title} {\enquote {\bibinfo {title} {Deterministic ising dynamics},}\ }\href@noop {} {\bibfield  {journal} {\bibinfo  {journal} {Annals of Physics}\ }\textbf {\bibinfo {volume} {167}},\ \bibinfo {pages} {62–72} (\bibinfo {year} {1986})}\BibitemShut {NoStop}%
\bibitem [{\citenamefont {Wetterich}(2018{\natexlab{a}})}]{CWIT}%
  \BibitemOpen
  \bibfield  {author} {\bibinfo {author} {\bibfnamefont {C.}~\bibnamefont {Wetterich}},\ }\bibfield  {title} {\enquote {\bibinfo {title} {{Information transport in classical statistical systems}},}\ }\href {\doibase 10.1016/j.nuclphysb.2017.12.008} {\bibfield  {journal} {\bibinfo  {journal} {Nucl. Phys. B}\ }\textbf {\bibinfo {volume} {927}},\ \bibinfo {pages} {35–96} (\bibinfo {year} {2018}{\natexlab{a}})},\ \Eprint {http://arxiv.org/abs/1611.04820} {arXiv:1611.04820} \BibitemShut {NoStop}%
\bibitem [{\citenamefont {Wetterich}(2018{\natexlab{b}})}]{CWQF}%
  \BibitemOpen
  \bibfield  {author} {\bibinfo {author} {\bibfnamefont {C.}~\bibnamefont {Wetterich}},\ }\bibfield  {title} {\enquote {\bibinfo {title} {{Quantum formalism for classical statistics}},}\ }\href {\doibase 10.1016/j.aop.2018.03.022} {\bibfield  {journal} {\bibinfo  {journal} {Annals Phys.}\ }\textbf {\bibinfo {volume} {393}},\ \bibinfo {pages} {1–70} (\bibinfo {year} {2018}{\natexlab{b}})},\ \Eprint {http://arxiv.org/abs/1706.01772} {arXiv:1706.01772} \BibitemShut {NoStop}%
\bibitem [{\citenamefont {Wetterich}(2020)}]{CWPW2020}%
  \BibitemOpen
  \bibfield  {author} {\bibinfo {author} {\bibfnamefont {C.}~\bibnamefont {Wetterich}},\ }\bibfield  {title} {\enquote {\bibinfo {title} {{The probabilistic world}},}\ }\href@noop {} {\  (\bibinfo {year} {2020})},\ \Eprint {http://arxiv.org/abs/2011.02867} {arXiv:2011.02867} \BibitemShut {NoStop}%
\bibitem [{\citenamefont {{'t Hooft}}(2014)}]{GTH}%
  \BibitemOpen
  \bibfield  {author} {\bibinfo {author} {\bibfnamefont {Gerard}\ \bibnamefont {{'t Hooft}}},\ }\bibfield  {title} {\enquote {\bibinfo {title} {{The Cellular Automaton Interpretation of Quantum Mechanics. A View on the Quantum Nature of our Universe, Compulsory or Impossible?}}}\ }\href@noop {} {\  (\bibinfo {year} {2014})},\ \Eprint {http://arxiv.org/abs/1405.1548} {arXiv:1405.1548} \BibitemShut {NoStop}%
\bibitem [{\citenamefont {Elze}(2014)}]{ELZE}%
  \BibitemOpen
  \bibfield  {author} {\bibinfo {author} {\bibfnamefont {Hans-Thomas}\ \bibnamefont {Elze}},\ }\bibfield  {title} {\enquote {\bibinfo {title} {{Quantumness of discrete Hamiltonian cellular automata}},}\ }\href {\doibase 10.1051/epjconf/20147802005} {\bibfield  {journal} {\bibinfo  {journal} {EPJ Web Conf.}\ }\textbf {\bibinfo {volume} {78}},\ \bibinfo {pages} {02005} (\bibinfo {year} {2014})},\ \Eprint {http://arxiv.org/abs/1407.2160} {arXiv:1407.2160} \BibitemShut {NoStop}%
\bibitem [{\citenamefont {{'t Hooft}}(2010)}]{HOOFT2}%
  \BibitemOpen
  \bibfield  {author} {\bibinfo {author} {\bibfnamefont {Gerard}\ \bibnamefont {{'t Hooft}}},\ }\bibfield  {title} {\enquote {\bibinfo {title} {{Classical cellular automata and quantum field theory}},}\ }\href {\doibase 10.1142/S0217751X10050469} {\bibfield  {journal} {\bibinfo  {journal} {Int. J. Mod. Phys. A}\ }\textbf {\bibinfo {volume} {25}},\ \bibinfo {pages} {4385–4396} (\bibinfo {year} {2010})}\BibitemShut {NoStop}%
\bibitem [{\citenamefont {{'t Hooft}}(2021)}]{HOOFT3}%
  \BibitemOpen
  \bibfield  {author} {\bibinfo {author} {\bibfnamefont {Gerard}\ \bibnamefont {{'t Hooft}}},\ }\bibfield  {title} {\enquote {\bibinfo {title} {{Fast Vacuum Fluctuations and the Emergence of Quantum Mechanics}},}\ }\href {\doibase 10.1007/s10701-021-00464-7} {\bibfield  {journal} {\bibinfo  {journal} {Found. Phys.}\ }\textbf {\bibinfo {volume} {51}},\ \bibinfo {pages} {63} (\bibinfo {year} {2021})},\ \Eprint {http://arxiv.org/abs/2010.02019} {arXiv:2010.02019} \BibitemShut {NoStop}%
\bibitem [{\citenamefont {Hooft}(2021)}]{HOOFT4}%
  \BibitemOpen
  \bibfield  {author} {\bibinfo {author} {\bibfnamefont {Gerard~t.}\ \bibnamefont {Hooft}},\ }\bibfield  {title} {\enquote {\bibinfo {title} {{Explicit construction of Local Hidden Variables for any quantum theory up to any desired accuracy}},}\ }\href@noop {} {\  (\bibinfo {year} {2021})},\ \Eprint {http://arxiv.org/abs/2103.04335} {arXiv:2103.04335} \BibitemShut {NoStop}%
\bibitem [{\citenamefont {Abdalla}\ \emph {et~al.}(1991)\citenamefont {Abdalla}, \citenamefont {Abdalla},\ and\ \citenamefont {Rothe}}]{AAR}%
  \BibitemOpen
  \bibfield  {author} {\bibinfo {author} {\bibfnamefont {E.}~\bibnamefont {Abdalla}}, \bibinfo {author} {\bibfnamefont {M.C.B.}\ \bibnamefont {Abdalla}}, \ and\ \bibinfo {author} {\bibfnamefont {K.D.}\ \bibnamefont {Rothe}},\ }\href@noop {} {\emph {\bibinfo {title} {{Nonperturbative methods in two-dimensional quantum field theory}}}}\ (\bibinfo {year} {1991})\BibitemShut {NoStop}%
\bibitem [{\citenamefont {Grinstein}\ \emph {et~al.}(1985)\citenamefont {Grinstein}, \citenamefont {Jayaprakash},\ and\ \citenamefont {He}}]{GJH}%
  \BibitemOpen
  \bibfield  {author} {\bibinfo {author} {\bibfnamefont {G.}~\bibnamefont {Grinstein}}, \bibinfo {author} {\bibfnamefont {C.}~\bibnamefont {Jayaprakash}}, \ and\ \bibinfo {author} {\bibfnamefont {Yu}~\bibnamefont {He}},\ }\bibfield  {title} {\enquote {\bibinfo {title} {Statistical {Mechanics} of {Probabilistic} {Cellular} {Automata}},}\ }\href {\doibase 10.1103/PhysRevLett.55.2527} {\bibfield  {journal} {\bibinfo  {journal} {Physical Review Letters}\ }\textbf {\bibinfo {volume} {55}},\ \bibinfo {pages} {2527–2530} (\bibinfo {year} {1985})},\ \bibinfo {note} {publisher: American Physical Society}\BibitemShut {NoStop}%
\bibitem [{\citenamefont {Lebowitz}\ \emph {et~al.}(1990)\citenamefont {Lebowitz}, \citenamefont {Maes},\ and\ \citenamefont {Speer}}]{LMS}%
  \BibitemOpen
  \bibfield  {author} {\bibinfo {author} {\bibfnamefont {Joel~L.}\ \bibnamefont {Lebowitz}}, \bibinfo {author} {\bibfnamefont {Christian}\ \bibnamefont {Maes}}, \ and\ \bibinfo {author} {\bibfnamefont {Eugene~R.}\ \bibnamefont {Speer}},\ }\bibfield  {title} {\enquote {\bibinfo {title} {Statistical mechanics of probabilistic cellular automata},}\ }\href {\doibase 10.1007/BF01015566} {\bibfield  {journal} {\bibinfo  {journal} {Journal of Statistical Physics}\ }\textbf {\bibinfo {volume} {59}},\ \bibinfo {pages} {117–170} (\bibinfo {year} {1990})}\BibitemShut {NoStop}%
\bibitem [{\citenamefont {Petersen}\ and\ \citenamefont {Alstrøm}(1997)}]{PA}%
  \BibitemOpen
  \bibfield  {author} {\bibinfo {author} {\bibfnamefont {Niels~K.}\ \bibnamefont {Petersen}}\ and\ \bibinfo {author} {\bibfnamefont {Preben}\ \bibnamefont {Alstrøm}},\ }\bibfield  {title} {\enquote {\bibinfo {title} {Phase transition in an elementary probabilistic cellular automaton},}\ }\href {\doibase 10.1016/S0378-4371(96)00410-4} {\bibfield  {journal} {\bibinfo  {journal} {Physica A: Statistical Mechanics and its Applications}\ }\textbf {\bibinfo {volume} {235}},\ \bibinfo {pages} {473–485} (\bibinfo {year} {1997})}\BibitemShut {NoStop}%
\bibitem [{\citenamefont {Fukś}(2003)}]{FU}%
  \BibitemOpen
  \bibfield  {author} {\bibinfo {author} {\bibfnamefont {Henryk}\ \bibnamefont {Fukś}},\ }\bibfield  {title} {\enquote {\bibinfo {title} {Probabilistic cellular automata with conserved quantities},}\ }\href {\doibase 10.1088/0951-7715/17/1/010} {\bibfield  {journal} {\bibinfo  {journal} {Nonlinearity}\ }\textbf {\bibinfo {volume} {17}},\ \bibinfo {pages} {159} (\bibinfo {year} {2003})}\BibitemShut {NoStop}%
\bibitem [{\citenamefont {Mairesse}\ and\ \citenamefont {Marcovici}(2014)}]{MM}%
  \BibitemOpen
  \bibfield  {author} {\bibinfo {author} {\bibfnamefont {Jean}\ \bibnamefont {Mairesse}}\ and\ \bibinfo {author} {\bibfnamefont {Irène}\ \bibnamefont {Marcovici}},\ }\bibfield  {title} {\enquote {\bibinfo {title} {Around probabilistic cellular automata},}\ }\href {\doibase 10.1016/j.tcs.2014.09.009} {\bibfield  {journal} {\bibinfo  {journal} {Theoretical Computer Science}\ }\bibinfo {series} {Non-uniform {Cellular} {Automata}},\ \textbf {\bibinfo {volume} {559}},\ \bibinfo {pages} {42–72} (\bibinfo {year} {2014})}\BibitemShut {NoStop}%
\bibitem [{\citenamefont {Ray}\ \emph {et~al.}(2024)\citenamefont {Ray}, \citenamefont {Laflamme},\ and\ \citenamefont {Kubica}}]{RLK}%
  \BibitemOpen
  \bibfield  {author} {\bibinfo {author} {\bibfnamefont {Annie}\ \bibnamefont {Ray}}, \bibinfo {author} {\bibfnamefont {Raymond}\ \bibnamefont {Laflamme}}, \ and\ \bibinfo {author} {\bibfnamefont {Aleksander}\ \bibnamefont {Kubica}},\ }\bibfield  {title} {\enquote {\bibinfo {title} {Protecting information via probabilistic cellular automata},}\ }\href {\doibase 10.1103/PhysRevE.109.044141} {\bibfield  {journal} {\bibinfo  {journal} {Physical Review E}\ }\textbf {\bibinfo {volume} {109}},\ \bibinfo {pages} {044141} (\bibinfo {year} {2024})},\ \bibinfo {note} {publisher: American Physical Society}\BibitemShut {NoStop}%
\bibitem [{\citenamefont {Verhagen}(1976)}]{VER}%
  \BibitemOpen
  \bibfield  {author} {\bibinfo {author} {\bibfnamefont {A.~M.~W.}\ \bibnamefont {Verhagen}},\ }\bibfield  {title} {\enquote {\bibinfo {title} {An exactly soluble case of the triangular ising model in a magnetic field},}\ }\href {\doibase 10.1007/BF01012878} {\bibfield  {journal} {\bibinfo  {journal} {Journal of Statistical Physics}\ }\textbf {\bibinfo {volume} {15}},\ \bibinfo {pages} {219–231} (\bibinfo {year} {1976})}\BibitemShut {NoStop}%
\bibitem [{\citenamefont {Peschel}\ and\ \citenamefont {Emery}(1981)}]{PE}%
  \BibitemOpen
  \bibfield  {author} {\bibinfo {author} {\bibfnamefont {I.}~\bibnamefont {Peschel}}\ and\ \bibinfo {author} {\bibfnamefont {V.~J.}\ \bibnamefont {Emery}},\ }\bibfield  {title} {\enquote {\bibinfo {title} {Calculation of spin correlations in two-dimensional {Ising} systems from one-dimensional kinetic models},}\ }\href {\doibase 10.1007/BF01297524} {\bibfield  {journal} {\bibinfo  {journal} {Zeitschrift für Physik B Condensed Matter}\ }\textbf {\bibinfo {volume} {43}},\ \bibinfo {pages} {241–249} (\bibinfo {year} {1981})}\BibitemShut {NoStop}%
\bibitem [{\citenamefont {Domany}\ and\ \citenamefont {Kinzel}(1984)}]{DK}%
  \BibitemOpen
  \bibfield  {author} {\bibinfo {author} {\bibfnamefont {Eytan}\ \bibnamefont {Domany}}\ and\ \bibinfo {author} {\bibfnamefont {Wolfgang}\ \bibnamefont {Kinzel}},\ }\bibfield  {title} {\enquote {\bibinfo {title} {Equivalence of {Cellular} {Automata} to {Ising} {Models} and {Directed} {Percolation}},}\ }\href {\doibase 10.1103/PhysRevLett.53.311} {\bibfield  {journal} {\bibinfo  {journal} {Physical Review Letters}\ }\textbf {\bibinfo {volume} {53}},\ \bibinfo {pages} {311–314} (\bibinfo {year} {1984})},\ \bibinfo {note} {publisher: American Physical Society}\BibitemShut {NoStop}%
\bibitem [{\citenamefont {Domany}(1984)}]{DOM}%
  \BibitemOpen
  \bibfield  {author} {\bibinfo {author} {\bibfnamefont {Eytan}\ \bibnamefont {Domany}},\ }\bibfield  {title} {\enquote {\bibinfo {title} {Exact {Results} for {Two}- and {Three}-{Dimensional} {Ising} and {Potts} {Models}},}\ }\href {\doibase 10.1103/PhysRevLett.52.871} {\bibfield  {journal} {\bibinfo  {journal} {Physical Review Letters}\ }\textbf {\bibinfo {volume} {52}},\ \bibinfo {pages} {871–874} (\bibinfo {year} {1984})},\ \bibinfo {note} {publisher: American Physical Society}\BibitemShut {NoStop}%
\bibitem [{\citenamefont {Rujàn}(1987)}]{RU}%
  \BibitemOpen
  \bibfield  {author} {\bibinfo {author} {\bibfnamefont {Pàl}\ \bibnamefont {Rujàn}},\ }\bibfield  {title} {\enquote {\bibinfo {title} {Cellular automata and statistical mechanical models},}\ }\href {\doibase 10.1007/BF01009958} {\bibfield  {journal} {\bibinfo  {journal} {Journal of Statistical Physics}\ }\textbf {\bibinfo {volume} {49}},\ \bibinfo {pages} {139–222} (\bibinfo {year} {1987})}\BibitemShut {NoStop}%
\bibitem [{\citenamefont {Georges}\ and\ \citenamefont {{Le Doussal}}(1989)}]{GD}%
  \BibitemOpen
  \bibfield  {author} {\bibinfo {author} {\bibfnamefont {Antoine}\ \bibnamefont {Georges}}\ and\ \bibinfo {author} {\bibfnamefont {Pierre}\ \bibnamefont {{Le Doussal}}},\ }\bibfield  {title} {\enquote {\bibinfo {title} {From equilibrium spin models to probabilistic cellular automata},}\ }\href {\doibase 10.1007/BF01019786} {\bibfield  {journal} {\bibinfo  {journal} {Journal of Statistical Physics}\ }\textbf {\bibinfo {volume} {54}},\ \bibinfo {pages} {1011–1064} (\bibinfo {year} {1989})}\BibitemShut {NoStop}%
\bibitem [{\citenamefont {Wetterich}(2010{\natexlab{a}})}]{CWFCS}%
  \BibitemOpen
  \bibfield  {author} {\bibinfo {author} {\bibfnamefont {C.}~\bibnamefont {Wetterich}},\ }\bibfield  {title} {\enquote {\bibinfo {title} {{Fermions from classical statistics}},}\ }\href {\doibase 10.1016/j.aop.2010.07.003} {\bibfield  {journal} {\bibinfo  {journal} {Annals Phys.}\ }\textbf {\bibinfo {volume} {325}},\ \bibinfo {pages} {2750–2786} (\bibinfo {year} {2010}{\natexlab{a}})},\ \Eprint {http://arxiv.org/abs/1006.4254} {arXiv:1006.4254} \BibitemShut {NoStop}%
\bibitem [{\citenamefont {Wetterich}(2017)}]{CWFGI}%
  \BibitemOpen
  \bibfield  {author} {\bibinfo {author} {\bibfnamefont {C.}~\bibnamefont {Wetterich}},\ }\bibfield  {title} {\enquote {\bibinfo {title} {{Fermions as generalized Ising models}},}\ }\href {\doibase 10.1016/j.nuclphysb.2017.02.012} {\bibfield  {journal} {\bibinfo  {journal} {Nucl. Phys. B}\ }\textbf {\bibinfo {volume} {917}},\ \bibinfo {pages} {241–271} (\bibinfo {year} {2017})},\ \Eprint {http://arxiv.org/abs/1612.06695} {arXiv:1612.06695} \BibitemShut {NoStop}%
\bibitem [{\citenamefont {Wetterich}(2021{\natexlab{a}})}]{CWPCA}%
  \BibitemOpen
  \bibfield  {author} {\bibinfo {author} {\bibfnamefont {C.}~\bibnamefont {Wetterich}},\ }\bibfield  {title} {\enquote {\bibinfo {title} {Probabilistic cellular automata for interacting fermionic quantum field theories},}\ }\href@noop {} {\bibfield  {journal} {\bibinfo  {journal} {Nuclear Physics B}\ }\textbf {\bibinfo {volume} {963}},\ \bibinfo {pages} {115296} (\bibinfo {year} {2021}{\natexlab{a}})},\ \Eprint {http://arxiv.org/abs/2007.06366} {arXiv:2007.06366} \BibitemShut {NoStop}%
\bibitem [{\citenamefont {Wetterich}(2021{\natexlab{b}})}]{CWFCB}%
  \BibitemOpen
  \bibfield  {author} {\bibinfo {author} {\bibfnamefont {Christof}\ \bibnamefont {Wetterich}},\ }\bibfield  {title} {\enquote {\bibinfo {title} {{Quantum fermions from classical bits}},}\ }\href@noop {} {\  (\bibinfo {year} {2021}{\natexlab{b}})},\ \Eprint {http://arxiv.org/abs/2106.15517} {arXiv:2106.15517} \BibitemShut {NoStop}%
\bibitem [{\citenamefont {Wetterich}(2021{\natexlab{c}})}]{CWNEW}%
  \BibitemOpen
  \bibfield  {author} {\bibinfo {author} {\bibfnamefont {C.}~\bibnamefont {Wetterich}},\ }\href@noop {} {\enquote {\bibinfo {title} {Fermionic quantum field theories as probabilistic cellular automata},}\ } (\bibinfo {year} {2021}{\natexlab{c}}),\ \Eprint {http://arxiv.org/abs/2111.06728} {arXiv:2111.06728 [hep-lat]} \BibitemShut {NoStop}%
\bibitem [{\citenamefont {Wetterich}(2022{\natexlab{a}})}]{CWFPPCA}%
  \BibitemOpen
  \bibfield  {author} {\bibinfo {author} {\bibfnamefont {C.}~\bibnamefont {Wetterich}},\ }\bibfield  {title} {\enquote {\bibinfo {title} {Fermion picture for cellular automata},}\ }\href {http://arxiv.org/abs/2203.14081} {\  (\bibinfo {year} {2022}{\natexlab{a}})},\ \Eprint {http://arxiv.org/abs/2203.14081} {arXiv:2203.14081} \BibitemShut {NoStop}%
\bibitem [{\citenamefont {Wetterich}(2022{\natexlab{b}})}]{CWPCAQP}%
  \BibitemOpen
  \bibfield  {author} {\bibinfo {author} {\bibfnamefont {C.}~\bibnamefont {Wetterich}},\ }\bibfield  {title} {\enquote {\bibinfo {title} {Probabilistic cellular automaton for quantum particle in a potential},}\ }\href@noop {} {\  (\bibinfo {year} {2022}{\natexlab{b}})},\ \Eprint {http://arxiv.org/abs/2211.17034} {arXiv:2211.17034 [quant-ph]} \BibitemShut {NoStop}%
\bibitem [{\citenamefont {Bell}(1964)}]{BELL}%
  \BibitemOpen
  \bibfield  {author} {\bibinfo {author} {\bibfnamefont {J.~S.}\ \bibnamefont {Bell}},\ }\bibfield  {title} {\enquote {\bibinfo {title} {{On the Einstein-Podolsky-Rosen paradox}},}\ }\href {\doibase 10.1103/PhysicsPhysiqueFizika.1.195} {\bibfield  {journal} {\bibinfo  {journal} {Physics Physique Fizika}\ }\textbf {\bibinfo {volume} {1}},\ \bibinfo {pages} {195–200} (\bibinfo {year} {1964})}\BibitemShut {NoStop}%
\bibitem [{\citenamefont {Clauser}\ \emph {et~al.}(1969)\citenamefont {Clauser}, \citenamefont {Horne}, \citenamefont {Shimony},\ and\ \citenamefont {Holt}}]{CHSH}%
  \BibitemOpen
  \bibfield  {author} {\bibinfo {author} {\bibfnamefont {John~F.}\ \bibnamefont {Clauser}}, \bibinfo {author} {\bibfnamefont {Michael~A.}\ \bibnamefont {Horne}}, \bibinfo {author} {\bibfnamefont {Abner}\ \bibnamefont {Shimony}}, \ and\ \bibinfo {author} {\bibfnamefont {Richard~A.}\ \bibnamefont {Holt}},\ }\bibfield  {title} {\enquote {\bibinfo {title} {{Proposed experiment to test local hidden variable theories}},}\ }\href {\doibase 10.1103/PhysRevLett.23.880} {\bibfield  {journal} {\bibinfo  {journal} {Phys. Rev. Lett.}\ }\textbf {\bibinfo {volume} {23}},\ \bibinfo {pages} {880–884} (\bibinfo {year} {1969})}\BibitemShut {NoStop}%
\bibitem [{\citenamefont {Wetterich}(2010{\natexlab{b}})}]{CWQPCS}%
  \BibitemOpen
  \bibfield  {author} {\bibinfo {author} {\bibfnamefont {C.}~\bibnamefont {Wetterich}},\ }\bibfield  {title} {\enquote {\bibinfo {title} {{Quantum particles from classical statistics}},}\ }\href {\doibase 10.1002/andp.201000088} {\bibfield  {journal} {\bibinfo  {journal} {Annalen Phys.}\ }\textbf {\bibinfo {volume} {522}},\ \bibinfo {pages} {807} (\bibinfo {year} {2010}{\natexlab{b}})},\ \Eprint {http://arxiv.org/abs/0904.3048} {arXiv:0904.3048} \BibitemShut {NoStop}%
\bibitem [{\citenamefont {Wetterich}(2011)}]{wetterich_classical_2011}%
  \BibitemOpen
  \bibfield  {author} {\bibinfo {author} {\bibfnamefont {C.}~\bibnamefont {Wetterich}},\ }\bibfield  {title} {\enquote {\bibinfo {title} {Classical probabilities for {Majorana} and {Weyl} spinors},}\ }\href {\doibase 10.1016/j.aop.2011.04.005} {\bibfield  {journal} {\bibinfo  {journal} {Annals of Physics}\ }\textbf {\bibinfo {volume} {326}},\ \bibinfo {pages} {2243–2293} (\bibinfo {year} {2011})},\ \bibinfo {note} {arXiv:1102.3586 [hep-lat, physics:hep-th, physics:quant-ph]}\BibitemShut {NoStop}%
\bibitem [{\citenamefont {Wetterich}(2010{\natexlab{c}})}]{CWQMCS}%
  \BibitemOpen
  \bibfield  {author} {\bibinfo {author} {\bibfnamefont {C.}~\bibnamefont {Wetterich}},\ }\bibfield  {title} {\enquote {\bibinfo {title} {{Quantum mechanics from classical statistics}},}\ }\href {\doibase 10.1016/j.aop.2009.12.006} {\bibfield  {journal} {\bibinfo  {journal} {Annals Phys.}\ }\textbf {\bibinfo {volume} {325}},\ \bibinfo {pages} {852} (\bibinfo {year} {2010}{\natexlab{c}})},\ \Eprint {http://arxiv.org/abs/0906.4919} {arXiv:0906.4919} \BibitemShut {NoStop}%
\bibitem [{\citenamefont {Wetterich}(2009)}]{CWEM}%
  \BibitemOpen
  \bibfield  {author} {\bibinfo {author} {\bibfnamefont {C.}~\bibnamefont {Wetterich}},\ }\bibfield  {title} {\enquote {\bibinfo {title} {{Emergence of quantum mechanics from classical statistics}},}\ }\href {\doibase 10.1088/1742-6596/174/1/012008} {\bibfield  {journal} {\bibinfo  {journal} {J. Phys. Conf. Ser.}\ }\textbf {\bibinfo {volume} {174}},\ \bibinfo {pages} {012008} (\bibinfo {year} {2009})},\ \Eprint {http://arxiv.org/abs/0811.0927} {arXiv:0811.0927} \BibitemShut {NoStop}%
\bibitem [{\citenamefont {Wetterich}(2012)}]{CWPT}%
  \BibitemOpen
  \bibfield  {author} {\bibinfo {author} {\bibfnamefont {C.}~\bibnamefont {Wetterich}},\ }\bibfield  {title} {\enquote {\bibinfo {title} {{Probabilistic Time}},}\ }\href {\doibase 10.1007/s10701-012-9675-3} {\bibfield  {journal} {\bibinfo  {journal} {Found. Phys.}\ }\textbf {\bibinfo {volume} {42}},\ \bibinfo {pages} {1384–1443} (\bibinfo {year} {2012})},\ \Eprint {http://arxiv.org/abs/1002.2593} {arXiv:1002.2593} \BibitemShut {NoStop}%
\bibitem [{\citenamefont {Xu}\ \emph {et~al.}(2013)\citenamefont {Xu}, \citenamefont {Shao},\ and\ \citenamefont {Tang}}]{xu_numerical_2013}%
  \BibitemOpen
  \bibfield  {author} {\bibinfo {author} {\bibfnamefont {Jian}\ \bibnamefont {Xu}}, \bibinfo {author} {\bibfnamefont {Sihong}\ \bibnamefont {Shao}}, \ and\ \bibinfo {author} {\bibfnamefont {Huazhong}\ \bibnamefont {Tang}},\ }\bibfield  {title} {\enquote {\bibinfo {title} {Numerical methods for nonlinear {Dirac} equation},}\ }\href {\doibase 10.1016/j.jcp.2013.03.031} {\bibfield  {journal} {\bibinfo  {journal} {Journal of Computational Physics}\ }\textbf {\bibinfo {volume} {245}},\ \bibinfo {pages} {131–149} (\bibinfo {year} {2013})}\BibitemShut {NoStop}%
\end{thebibliography}%

\end{document}